# Diffraction of Light from Optical Fourier Surfaces

Yannik M. Glauser,[†] J. J. Erik Maris,[†] Raphael Brechbühler,[†] Juri G. Crimmann,[†] Valentina G. De Rosa,[†] Daniel Petter,[†] Gabriel Nagamine,[†] Nolan Lassaline,[†,§] and David J. Norris[*,†]

[†]Optical Materials Engineering Laboratory, Department of Mechanical and Process Engineering, ETH Zurich, 8092 Zurich, Switzerland

[§]Department of Physics, Technical University of Denmark, 2800 Kongens Lyngby, Denmark

ABSTRACT. Diffractive surfaces shape optical wavefronts for applications in spectroscopy, high-speed communication, and imaging. The performance of these structures is primarily determined by how precisely they can be patterned. Fabrication constraints commonly lead to square-shaped, "binary" profiles that contain unwanted spatial frequencies that contaminate the diffraction. Recently, "wavy" surfaces (known as optical Fourier surfaces, OFSs) have been introduced that include only the desired spatial frequencies. However, the optical performance and reliability of these structures have not yet been experimentally tested with respect to models and simulations. Such a quantitative investigation could also provide previously unobtainable information about the diffraction process from the most fundamental diffractive surfaces—sinusoidally pure profiles. Here, we produce and study two classes of reflective OFSs: (i) single-sinusoidal profiles of varying depth and (ii) double-sinusoidal profiles with varying relative phase. After refining our fabrication procedure to obtain larger and deeper OFSs at higher yields, we find that the measured optical responses from our OFSs agree quantitatively with full electrodynamic simulations. In contrast, our measurements diverge from analytical scalar diffraction models routinely used by researchers to describe diffraction. Overall, our results confirm that OFSs provide a precise and powerful platform for Fourier-spectrum engineering, satisfying the growing demand for intricately patterned interfaces for applications in holography, augmented reality, and optical computing.

**INTRODUCTION**

Many photonic devices require the propagation of light to be precisely controlled. For over a century, scientists have addressed this challenge by exploiting state-of-the-art manufacturing techniques to structure optical materials to manipulate light through diffraction. For example, the use of periodically patterned surfaces—diffraction gratings—is well-established in the fields of spectroscopy,[1-3] high-speed communication,[4-6] and imaging.[7-9] Beyond periodic gratings, more sophisticated diffractive surfaces are driving advances in modern technologies such as holography,[10-12] augmented reality,[13,14] and optical computing.[15,16] Recent developments have depended on the precision with which interfaces can be patterned, where present-day variants utilize nanoscale surface structures with feature sizes similar to, or even smaller than, the wavelength of light. In particular, arrays of subwavelength nanoantennas have been developed to tailor the optical wavefront.[17-20] While such interfaces, known as metasurfaces, offer great versatility for manipulating light, more conventional diffractive structures still represent an important platform for photonic applications that continues to advance.[1-16]

Through lithographic patterning, diffractive surfaces can be engineered to manipulate light by altering the amplitude,[21-23] phase,[22-26] polarization,[27-29] and wavelength[30] of an impinging electromagnetic field. Phase gratings, which are periodic structures that modulate the phase of light, are commonly employed due to their high diffraction efficiency and straightforward integration within photonic devices.[5,22,23] Furthermore, phase gratings are often fabricated by structuring a single optical surface for use in transmission or reflection. In this context, they are frequently referred to as surface relief gratings. They offer a simple yet powerful approach for the design, fabrication, and use of diffractive elements in optical technologies.[31]

The physics of such elements can be described within the framework of Fourier optics,[22] where scalar diffraction theory is used to model complex optical systems with analytical formulas. This approach has been applied as a fast and inexpensive method to describe diffraction from many structures.[23,32,33] However, more rigorous treatments (*e.g.*, numerical scalar models, vectorial models, or even full electrodynamic simulations) are often required to accurately quantify how light interacts with the nanostructured environment.[32,34-40] Unfortunately, rigorous numerical simulations of the electromagnetic fields are frequently computationally prohibitive.

The compromise between speed, simplicity, and accuracy of the theoretical treatment is further complicated by the design–fabrication mismatch that exists between the desired structures and what is possible to produce. For example, due to fabrication constraints, most diffractive structures for visible or near-infrared light are restricted to square-shaped, "binary" surface profiles with only two depth levels.[26] This often leads to limitations in the design and implementation of diffractive optics. Consequently, despite the long history of diffractive surfaces, quantitative investigations of the most fundamental phase gratings—sinusoidally pure surface profiles—have not been possible. To our knowledge, no detailed comparisons exist for such systems between analytical models, numerical simulations, and experimental measurements.

Recently, optical Fourier surfaces (OFSs)[41] have been introduced as a class of wavy diffractive elements that allow mathematically precise topographies to be fabricated on the nanoscale. In contrast to diffractive surfaces produced *via* conventional grayscale lithographic methods, such as interference lithography,[42-47] OFSs are produced *via* thermal scanning-probe lithography (TSPL).[48,49] This leads to surface profiles with higher resolution and greater topographical precision, which is useful for Fourier-spectrum engineering.[41,50] Consequently, OFSs present a potential route to eliminate the design–fabrication mismatch often associated with phase gratings. They offer a well-controlled and efficient diffraction process based on simple physical insights from Fourier optics. However, the promise of OFSs in terms of their optical performance and reliability must be experimentally tested with respect to models and simulations. Such a quantitative investigation could also provide previously unobtainable information about the diffraction process from sinusoidally pure surface profiles.

In this work, we investigate reflective OFSs by comparing their measured diffraction behavior with scalar diffraction models and finite-difference time-domain (FDTD) simulations. We confirm that our fabricated structures provide sufficient quality that their optical responses match expectations from FDTD simulations. To obtain these results, we first improved the fabrication procedure from our previous work.[41] This allowed us to produce diffractive surfaces that are ~4× larger in area and up to

~50% deeper while exploiting higher sample yields. Using these capabilities, we then created a series of single- and double-sinusoidal profiles of varying depth and relative phase, respectively, and examined their optical responses as a function of wavelength. Single-sinusoidal OFSs represent the most fundamental diffractive surface profile, while double-sinusoidal OFSs allow us to test the superposition of multiple spatial frequencies. We observe that even for these elementary surface structures, common scalar diffraction models become inapplicable to frequently exploited diffraction regimes, *e.g.*, for large diffraction angles and deep structures, leading to intuition about the diffraction process. In such cases, researchers should be aware of the limitations of these analytical models and consider numerical simulations if computationally feasible. More generally, our findings demonstrate that OFSs are sufficiently precise to resolve the design–fabrication mismatch and provide a powerful platform for Fourier-spectrum engineering of diffractive surfaces. The need for such control has increased due to the recent interest in intricately patterned interfaces for applications in areas such as optical neural networks.[15,16]

**RESULTS AND DISCUSSION**

**Diffraction from Optical Fourier Surfaces.** OFSs are patterned surfaces that exploit wavy profiles to precisely diffract light. Figure 1a shows our fabrication process for silver (Ag) OFSs. Importantly, we utilized poly(phthalaldehyde) (PPA) as a thermally sensitive resist to improve sample area, depth, and yield compared to our previous results.[41] After patterning a film of PPA *via* TSPL,[48,49] Ag is thermally evaporated[51] and template stripped[52] to obtain the final structure on a glass substrate (see Methods for details). Figure 1b displays a schematic of the diffraction process. In this specific case, an input beam illuminates the OFS at normal incidence. The light is then reflected by the OFS and diffracted into the corresponding diffraction orders. Depending on the orientation of the electric field with respect to the modulation direction of the surface, the light is *s*- or *p*-polarized, as indicated in Figure 1b,c.

The surface profile of an OFS directly affects the phase of the wavefront by locally modulating the path length of the light and introducing optical path differences. These phase modulations at the surface plane appear as diffraction orders in a Fourier plane of our imaging system, as schematically illustrated in Figure S1 in the Supporting Information. A sinusoidal phase grating with period $\Lambda$ provides in-plane momentum $g = 2\pi/\Lambda$ to the incoming light, which has a free-space wavevector $k_0 = 2\pi/\lambda$ with $\lambda$ the optical wavelength in vacuum. The grating momentum $g$ represents the spatial frequency of the surface profile and corresponds to the lateral shift of the diffraction orders in Fourier space (Figure 1c). The diffraction orders are named according to the integer number of grating momenta added or subtracted. Diffraction orders with a total in-plane wavevector $\mathbf{k}_\parallel = (k_x, k_y)$ with $|\mathbf{k}_\parallel| \leq k_0$, represent propagating modes. They can be detected in the Fourier plane of our imaging system if the ratio $|\mathbf{k}_\parallel|/k_0$ is smaller than or equal to the numerical aperture (NA) of the system. Diffraction orders with a larger total in-plane wavevector with $|\mathbf{k}_\parallel| > k_0$ define evanescent modes that do not propagate away from the surface, which, therefore, cannot be measured in Fourier space.

Two specific designs of OFSs are studied in this work. First, we consider surfaces incorporating only a single sinusoid, for which the period $\Lambda$ is fixed (1000 nm) while the amplitude $A$ is varied between 0 and 150 nm (see schematic in Figure 1d). We exploit these simple OFSs to understand diffraction on a fundamental level by providing information about the diffraction efficiencies as a function of the optical wavelength and pattern amplitude. Second, we consider OFSs with two superposed sinusoids that have a specific relative phase $\varphi$. This phase is varied between 0 and $2\pi$, while the sinusoidal periods ($\Lambda_1 = 1000$ nm and $\Lambda_2 = 500$ nm) and amplitudes ($A_1 = 90$ nm and $A_2 = 45$ nm) are fixed. The second harmonic $g_2$ has half the period and half the amplitude compared to the fundamental sinusoid $g_1$ (see schematic in Figure 1e and Table S1 in the Supporting Information). By adding just one additional sinusoid to the system, more complex diffraction phenomena begin to emerge. Depending on the relative phase $\varphi$, the mirror symmetry of the surface profile with respect to the *yz* plane (Figure 1b) can be broken. In this case, light is preferably diffracted into one specific diffraction order, leading to an asymmetric diffraction process. This effect becomes important for more complicated directional structures such as blazed or apodized gratings.[23,53-60]

To investigate such surfaces, analytical scalar diffraction models have been reported.[22,23,32,61] While the underlying scalar diffraction theory can accurately describe light propagation in such systems, the analytical models based on this theory apply simplifying assumptions. Specifically, they primarily differ in two aspects: (i) the method of converting the surface profile into an optical path difference (*e.g.*, using the paraxial or non-paraxial approach) and (ii) the procedure for normalizing the evanescent diffraction orders (see Figure S2 and Section S1 in the Supporting Information). In this work, we analyze and discuss four models: (1) the paraxial approach with renormalization, (2) the paraxial approach without renormalization, (3) the non-paraxial approach with renormalization, and (4) Model 3 corrected for the diffracted radiance (see Methods and Supporting Information). Among these, we anticipated that the most accurate and physically intuitive model for our geometries was Model 3. This model accounts for large diffraction angles (non-paraxial approximation) and renormalizes diffraction efficiencies when diffraction orders become evanescent.[23,32] It is discussed here in the main text, while the other models are presented in Figures S3–S4 and Section S1 in the Supporting Information.

Figure 1d,e applies Model 3 to calculate the expected trends for our two different classes of OFSs. Figure 1d plots the expected total 1st-order diffraction efficiency (+1st and −1st) as a function of the amplitude $A$, which is half of the OFS depth. A discontinuity can be observed at $\lambda = 500$ nm, where the 2nd orders become evanescent ($2g > k_0$), and their intensities are redistributed to the remaining propagating orders (known commonly as a Rayleigh anomaly).[23,32] The diffraction behavior also changes with increasing amplitude. The maximum diffraction efficiency increases initially before saturating, and the corresponding wavelength for maximum efficiency shifts to the red.

Figure 1e depicts the expected diffraction efficiency for the +1st order as a function of the relative phase $\varphi$ (0 to $\pi$). The +1st order becomes less efficient with increasing $\varphi$, sending more light to the −1st order. If $\varphi$ were further increased from $\pi$ to $2\pi$, the opposite trend would be observed. Intuitively, this can be anticipated by considering the limiting cases ($\varphi = 0$ and $\pi$; Figure 1e), which represent blazed gratings oriented toward the right or left side, respectively.

**Surface-Topography Characterization.** Figure 2a,c shows scanning-electron microscopy (SEM) images of a single-sinusoidal OFS in Ag ($A = 105$ nm and $\Lambda = 1000$ nm) with a lateral size of 40 $\mu$m × 40 $\mu$m (which is ~4× larger than our previous OFSs[41]). Figure 2b displays the measured topography (circles) and fitted sinusoid (line) of the patterned PPA from which this Ag OFS was obtained. The TSPL tool provides such topography data during patterning. Extensive measurements have shown that such surface profiles are maintained (see Methods) after the transfer into Ag. In our discussions below, we utilize our PPA data to characterize the OFS topographies. Note that, due to template stripping, the initial structure in PPA is inverted and mirrored with respect to the final OFS in Ag. Figure 2e–h shows SEM images and topography data for two of our double-sinusoidal OFSs with $\varphi = 0$ and $\varphi = \pi$, respectively.

Figure 2d summarizes the topography characterization for our single-sinusoidal OFSs: 30 structures with amplitudes from $A = 5$ nm to $A = 150$ nm (which is ~50% deeper than in our previous work[41]). These structures were fabricated together on one substrate (see dark-field images in Figure S5a in the Supporting Information). Each is plotted as a purple dot according to its designed and fitted amplitude on the $x$ and $y$ axes, respectively. A good agreement is observed between the design and measurements. The root-mean-square error (RMSE) between the fitted sinusoid and measured profile is also shown (green dots) for the structured region. Here, the RMSE represents both surface roughness and systematic errors in the surface profile that occur due to fabrication imperfections. We see that the RMSE increases linearly with amplitude up to $A = 125$ nm and then grows faster for deeper OFSs. An advantage of TSPL is that RMSEs of a few nanometers are reliably obtained.

Similarly, Figure 2i shows the topography characterization for our double-sinusoidal OFSs: 12 structures that were fabricated together on one substrate (Figures S5b and S6 in the Supporting Information). The fitted relative phases show good agreement with the desired values (purple dots). Indeed, a second advantage of TSPL is the ability to precisely superpose sinusoids with a specific relative phase. The RMSE between the double-sinusoidal fit and the measured profile is indicated by green dots. Except for three outliers, the RMSE is ~5 nm. This is consistent with our single-sinusoidal OFSs of comparable depth in Figure 2d.

**Diffraction Measurements.** To optically measure the diffraction behavior of the individual OFSs, we used a home-built Fourier microscope (Figure 3a).[62] A supercontinuum laser source with a tunable band-pass filter (filterbox) allows input wavelengths from 450 to 745 nm. The light is focused on the back focal plane of the objective (NA of 0.8) to illuminate the sample with a plane wave at normal incidence (approximated as a large-diameter Gaussian beam). Optics then relay the reflected diffraction output from the back focal plane of the objective (Fourier space) to a camera. In between, a circular iris is placed in an image plane (real space) to collect only light that is diffracted from the patterned OFS. Orthogonal linear polarizers in the input and output paths are used to independently analyze *s*- and *p*-polarized light. Below we discuss measurements with *s*-polarized light that avoid coupling to surface plasmon polaritons. (Measurements for *p*-polarization are shown in the Supporting Information.)

The diffraction efficiencies for our OFSs were determined by reflectivity measurements. The diffraction output from each sample was normalized by a reference measurement from a flat Ag surface. We note that this ignores the absorption from the Ag, which is a few percent.[41] Figure 3b,c shows the reference and sample measurements for 532 nm illumination. The black circular area represents propagating modes in Fourier space (*i.e.*, inside the light cone). The area that can be practically measured with our objective is indicated by a gray dashed circle. While the flat Ag shows only a $0^{\text{th}}$-order diffraction spot (Figure 3b), the sample (in this case, the OFS shown in Figure 2a–c) reveals two additional features, representing the $-1^{\text{st}}$ and $+1^{\text{st}}$ orders (Figure 3c). The width of the diffraction peaks is connected to the finite size of the OFS. Because the diffraction spots are distributed along $k_x$, we only measured signals between the green vertical lines in Figure 3b,c (from −0.2 to 0.2 $k_y/k_0$) to minimize the size of our datasets.

To evaluate diffraction efficiencies as a function of wavelength, we first integrated the counts along $k_y$ for both the reference and the sample at a specific $\lambda$ (Figure 3d). The collected counts were then integrated along $k_x$ (±0.2 $k_x/k_0$ from the expected position for each order; see pink regions in Figure 3d). We then repeated this procedure for each wavelength from 450 to 745 nm. Figure 3e depicts the wavelength-dependent diffraction signal for the $-1^{\text{st}}$, $0^{\text{th}}$, and $+1^{\text{st}}$ orders in Fourier space. (The expected positions of the $2^{\text{nd}}$-order diffraction features, which lie outside our measurable region in $k$ space, are indicated with solid white lines.) For each $\lambda$, the total collected counts per diffraction order were divided by the total reference counts to obtain the corresponding diffraction efficiency. A similar dataset was measured for each of our OFSs.

**Single-Sinusoidal OFSs as a Function of Profile Depth.** We first applied the above procedure to single-sinusoidal OFSs with varying amplitudes. Figure 4a–c shows the total $1^{st}$-order diffraction efficiency for different OFSs for experiment, simulation, and model, respectively. (Complementary data on the $0^{th}$- and total $2^{nd}$-order diffraction efficiencies are shown in Figure S7 in the Supporting Information.) The fitted amplitudes from the topography measurements (see Figure 2d) are used. In Figure 4a, each pixel corresponds to an individual measurement for a specific OFS and wavelength. The simulation data (Figure 4b) is based on FDTD simulations performed with Lumerical. Figure 4c plots expectations from the scalar diffraction model applied in Figure 1d,e (Model 3). The model and simulation data are interpolated at the points of the experimental data to facilitate direct comparisons.

In general, the data in Figure 4a–c for all approaches (experiment, simulation, and model) appear very similar. Our ability to directly compare the expectations from simulations and the model with such detailed experiments is only possible due to the precision of the OFSs. All three plots reveal the same wavelength- and amplitude-dependent trends. The diffraction efficiency at a fixed wavelength goes through a maximum with increasing amplitude. At longer wavelengths, this maximum occurs for deeper structures. In addition, the Rayleigh anomaly is clearly visible at $\lambda = 500$ nm.

However, differences do exist between simulation, model, and experiment, providing quantitative information about the diffraction process. Figure 4e compares simulation and experiment. In most of this plot, negligible differences are observed, confirming the quality of our OFSs. Only for our deepest structures and close to the Rayleigh anomaly do we observe discrepancies (up to ~15%). Because the simulation assumes an infinite structure, it predicts a more distinct Rayleigh anomaly compared to experiment, explaining differences in this region. The remaining deviations presumably originate in experiment, as we estimated the error of our simulations to be below 1% (based on convergence tests). Measurements from deeper structures can be affected by their higher RMSE (see Figure 2d) and PPA residues (see Methods and Supporting Information). Alternatively, our experiments scan a broad wavelength range, which leads to unavoidable chromatic aberrations. The impact of such aberrations can be estimated with Figure 4d, which shows the measured difference in efficiency between the −1st and +1st order. While theoretically, these efficiencies should be equal, we observe differences of up to ~5% at the top left of the plot. Thus, we assume that the structural imperfections in the deepest OFSs lead to experimental errors of ~10%.

Figure 4f shows larger differences between the model and experiment (from around −10 to 25%). While the model underestimates the experimental efficiencies for large amplitudes and short wavelengths (top left corner of the plot), it significantly overestimates the efficiencies for large amplitudes and long wavelengths (top right corner of the plot). These findings clearly show that even the most accurate scalar diffraction model that we investigated (Model 3) is insufficient despite its inclusion of non-paraxial effects and Rayleigh anomalies. (The efficiency data for all four tested models are shown in Figure S3 in the Supporting Information.) Indeed, Figure 4f can serve as a quantitative map for expected errors between experiment and scalar diffraction models for diffractive surfaces of a given depth and period.

The inconsistencies between the models and experiment are mainly caused by the underlying simplifications. In general, shallower structures (with smaller amplitudes $A$) that diffract at smaller angles (for shorter wavelengths $\lambda$ or longer periods $\Lambda$) can be well described. This is because most models exploit the paraxial approximation, which applies in these limits. However, even non-paraxial models like Model 3 fail for deeper structures and larger diffraction angles. In these limits, the approximations utilized in Fourier optics (*e.g.*, the Fraunhofer approximation or the more general Fresnel approximation) become invalid. From a more intuitive perspective, the classical ray picture used to calculate optical path differences breaks down (see Figure S2 in the Supporting Information).

**Double-Sinusoidal OFSs as a Function of Relative Phase.** For many applications where diffraction is used in free space or for in- and out-coupling of light to guided modes (*e.g.*, in grating couplers for integrated photonics), light should be diffracted asymmetrically in a specific direction. For the second measurement series, we therefore focused on OFSs with two superposed sinusoids, serving as a simple system that produces asymmetric diffraction.

Figure 5a–c shows the +1st-order diffraction efficiencies for experiment, simulation, and model (Model 3), respectively, analogous to Figure 4a–c. (Similar comparisons for the 0th- and −1st-order diffraction efficiencies are plotted in Figure S9 in the Supporting Information.) The relative phase $\varphi$ is varied from 0 to $2\pi$ (in steps of $\pi/6$, previously fitted in Figure 2i), and the wavelength $\lambda$ is swept from 450 to 745 nm. All three approaches exhibit similar trends. The +1st-order diffraction efficiency is higher toward $\varphi = 0$ and lower near $\varphi = \pi$, and the Rayleigh anomaly can be observed at $\lambda = 500$ nm. The efficiency is generally highest for wavelengths near the Rayleigh anomaly, peaking on the long-wavelength side. For structures with relative phase of $\varphi \approx 0$, the efficiency then decreases with increasing wavelength. This observation contrasts with the results from the single-sinusoidal OFSs of comparable depth ($A = 135$ nm, Figure 4a–c), which exhibit efficiency maxima at longer wavelengths.

To probe how the light is diffracted asymmetrically, Figure 5d shows the experimentally measured efficiency ratio between +1st- and −1st-order diffraction on a logarithmic scale. Near the limiting cases ($\varphi \approx 0$ and $\pi$), 18 and 24 times more light is diffracted to the +1st or −1st order, respectively. The maximum diffraction asymmetries occur on the short-wavelength side of the Rayleigh anomaly; the asymmetry ratio then decreases with increasing wavelength. These findings indicate that the diffraction process can become highly directional by adding just one additional sinusoid to our OFSs.

For a quantitative analysis of the +1st-order diffraction results in Figure 5a–c, we plotted the efficiency differences for the simulation (Figure 5e) and the model (Figure 5f), both with respect to the experiment. Between the simulation and experiment, the discrepancies lie below ~5% for most of the plot (Figure 5e), confirming again the quality of our OFSs. Slightly higher differences occur only around the Rayleigh anomaly, particularly on the short-wavelength side. This is likely due to the more distinct Rayleigh anomaly predicted by the simulations, as discussed above. The discrepancies between the model and experiment (Figure 5f) are more significant (from around −15 to 30%). The applied model (Model 3) overestimates and underestimates the efficiency in regions with strong and weak +1st-order diffraction, respectively. (Figure S4 in the Supporting Information shows results for all four investigated models.) Especially toward longer wavelengths (*i.e.*, larger diffraction angles), the model becomes increasingly unreliable. This observation follows the findings for single-sinusoidal OFSs. Because the structures analyzed in Figure 5 have fixed, relatively deep amplitudes ($A_1$ = 90 nm and $A_2$ = 45 nm), one would expect significant discrepancies based on Figure 4f. As discussed above, scalar diffraction models fail to properly describe deeper structures (with larger amplitudes $A$) that diffract at larger angles (for longer wavelengths $\lambda$ or shorter periods $\Lambda$). More generally, these observations indicate that the additional sinusoid in the OFS profile leads to increased discrepancies between the model and experiment. For example, the plot for the model (Figure 5c and Figure S4 in the Supporting Information) is symmetric around $\varphi = \pi$. Such a symmetry is not observed in experiment and simulation. The symmetry in the model occurs due to the form of the analytical diffraction-efficiency formula (see eq S33 in the Supporting Information). However, the actual structures do not exhibit geometrical symmetries around $\varphi = \pi$.

The observed unreliability of the scalar diffraction models is important to appreciate when dealing with more sophisticated structures, consisting of numerous spatial Fourier components. For example, in the design of advanced diffractive devices,[12,16] where paraxial scalar diffraction models are commonly applied, phase maps ranging from 0 to $2\pi$ are typically converted directly to surface profiles using simple optical path differences (on the order of $\lambda$). This approach can be improved by implementing non-paraxial corrections. However, as we have shown, these strategies can also become problematic in specific diffraction regimes. Researchers should be aware of these limitations, and more rigorous numerical simulations may be required for the design and validation of complicated diffractive surfaces.

**CONCLUSIONS**

We have presented a comprehensive study of diffraction from optical Fourier surfaces. Two classes of reflective OFSs, representing the most fundamental surface profiles (a single sinusoid and two superposed sinusoids), have been fabricated, topographically characterized, and optically measured to study the relationship between the surface profile and diffraction. We examined different scalar diffraction models and performed numerical FDTD simulations for comparison with measurements. Our results show that the diffraction from OFSs matches well with expectations from simulations, bridging the gap between theory and experiment and providing intuition for the design and analysis of next-generation diffractive elements. Interestingly, our findings also demonstrate that common scalar diffraction models become inapplicable to frequently exploited diffraction regimes, *e.g.*, for large diffraction angles and deep structures. Therefore, numerical simulations should be considered, especially for diffractive surfaces with profiles that contain many Fourier components.

These results were only possible due to the capabilities of OFSs to precisely incorporate desired spatial frequencies in a diffractive interface. We exploited an improved fabrication procedure that enabled larger and deeper structures with higher sample yields. Based on these advancements, the OFS platform now provides a route to perform fundamental optical experiments based on intuitive design rules and analytical mathematics from Fourier optics. The ability of our OFSs to match electromagnetic simulations also confirms their promise for use in applications such as holography, integrated photonics, and analogue optical computing. In contrast to conventional binary diffractive structures, the OFS platform offers a route to satisfy the increasingly challenging requirements for diffractive surfaces in these important technological areas.

**METHODS**

**Scalar Diffraction Models.** The four scalar diffraction models considered in this work are discussed in Section S1 in the Supporting Information. These models were implemented in Matlab (Release 2022b) to calculate the diffraction efficiencies for the different OFSs. Due to the limited size of the OFSs, the modelled diffraction outputs were broadened by convolution with an airy disc function. This represents the Fourier transform of a circular aperture of lateral size 40 $\mu$m × 40 $\mu$m, accounting for the iris in the collection path of the optical setup.

**Design of Fourier Surfaces.** The designs for our OFSs are based on analytical functions. The feasible structures are limited by fabrication constraints (*e.g.*, due to the shape of the TSPL tip[63]) in terms of lateral size, depth, gradient, and curvature. The lateral size is restricted by the scan range of the piezo stage (~50 $\mu$m). Our measurements show that the surface topography cannot contain any slopes larger than ~1. The design parameters for all of our OFSs are summarized in Table S1 in the Supporting Information. The OFSs were constructed such that the 1st-order diffraction was directly observable, while the 2nd-order diffraction could be indirectly measured through the 0th- and 1st-order diffraction (Rayleigh anomaly).

**Bitmap Generation.** The analytical design function was mapped onto a two-dimensional (2D) pixel grid with a lateral size of 40 $\mu$m × 40 $\mu$m. The pixel size was set to 20 nm × 20 nm. The values of the function were discretized into 256 depth levels (8-bit precision). This led to an 8-bit grayscale image in the form of a bitmap (.bmp file format). The bitmap generation was performed with Matlab (Release 2022b). To account for the process of template stripping, the final bitmaps were inverted in the $z$ direction and mirrored along the $yz$ plane (see Figure 1b).

**Thermal Scanning-Probe Lithography (TSPL).** Compared to our previous work on OFS fabrication,[41] in which we used PMMA/MA [poly(methyl methacrylate-co-methacrylic acid)] or CSAR [poly($\alpha$-methylstyrene-co-methyl chloroacrylate)] as the thermal resist, here we exploited PPA. This led to significantly less tip contamination and subsequently higher yields.

For sample preparation, a 1-mm-thick, 2-inch-diameter Si (100) wafer (Silicon Materials) was cleaned by oxygen plasma (GIGAbatch, PVA TePla) at 600 W for 2 min. The wafer was spin-coated with 300 $\mu$L of 12 wt% PPA (Allresist) dissolved in anisole (AR 600-02, Allresist). For spin-coating, we used a two-step recipe: (i) 5 s at 500 rpm with a ramp of 500 rpm/s and (ii) 40 s at 2000 rpm with 2000 rpm/s. This led to a PPA film thickness between 350 to 400 nm, which was confirmed with a profilometer (Dektak XT, Bruker).

The designed bitmaps were loaded into the TSPL tool (NanoFrazor Explore, Heidelberg Instruments), and the specific depth range for the structures was selected. A heated cantilever with a hot tip was scanned across the surface to pattern the PPA by sublimation. The depth of the pattern at each pixel was adjusted by a tunable downward force which resulted from an electrostatic potential applied between the cantilever and substrate. The TSPL tool also measures the surface topography simultaneously while writing the corresponding structures.

**Evaporation and Template Stripping.** An optically thick Ag film (>500 nm) was thermally evaporated (Nano 36, Kurt J. Lesker) on top of the patterned PPA layer. 1/4-inch-diameter × 1/4-inch-long Ag pellets (99.999%, Kurt J. Lesker) were used to deposit at a rate of 25 Å/s at a pressure of ~3 × 10$^{-7}$ mbar. After evaporation, a 1-mm-thick glass microscope slide (Paul Marienfeld) was affixed to the top of the Ag layer using an ultraviolet-curable epoxy (OG142-95, Epoxy Technology). To reduce residual PPA on the final structure, it is important to let the glass slide settle on the epoxy for ~5 min before then illuminating with ultraviolet light for 2 h. Afterwards, the glass/epoxy/Ag stack was stripped off the PPA with a razor blade to reveal the final surface structure in Ag, inverted and mirrored with respect to the original structure in PPA.

**Characterization of Surface Topography.** The main topography characterization was based on the PPA topography data from the TSPL tool. This information was analyzed in Matlab (Release 2022b). First, measurements from flat reference regions, which surrounded the patterned areas, were used to perform a 2D planar fit to level the measured topography. After this correction, the OFS data were subsequently fitted to either a single- or double-sinusoidal function. This produced the fit parameters and the RMSEs shown in Figure 2d,i. While $\Lambda$ was used as a fitting parameter in this procedure, the fit differed from the design period by 0.07% (averaged over all structures). The fitted amplitudes $A$ for the single-sinusoidal structures are shown in Figure 2d. They differed from the targeted amplitudes by less than 1%. For the double-sinusoidal OFSs, the fitted amplitudes differed from the targets by less than 2%.

The topography of the PPA structures was taken directly to represent that of the corresponding Ag structures. This was based on our assumption that the transfer between PPA and Ag did not noticeably affect the geometric properties of our patterns. This was validated by measuring the central region (10 $\mu$m × 10 $\mu$m) of a single-sinusoidal Ag OFS (shown in Figure 2a) with atomic force microscopy (AFM; MFP-3D Origin AFM, Oxford Instruments). By comparing the AFM data with the TSPL topography for the corresponding PPA structure, we observed a difference of ~1% in $\Lambda$. A similar deviation was previously reported[41] and was attributed to a consistent distance miscalibration of the TSPL tool, which could be corrected if desired. The TSPL tool provided better topographical data for the comparably large OFSs than the commercial AFMs to which we had access. Therefore, we relied on the PPA data for surface characterization.

Minor residues of PPA could be observed on ~20% of our Ag OFSs. They occurred especially for asymmetric structures with steep gradients in the surface profile. An SEM image showing the worst case is shown in Figure S10 in the Supporting Information. While the position of the diffraction spot in Fourier space should not be affected by such point defects, they may lead to slightly reduced diffraction efficiencies.

**Optical Measurements.** A supercontinuum laser (SuperK Fianium, NKT Photonics) with a tunable band-pass filter (LLTF Contrast, NKT Photonics) was used to sweep through a range of wavelengths from 450 to 745 nm with a linewidth of ~1.5 nm. After the laser light passed through an optical fiber (A502-010-110, NKT Photonics), a collimation lens (Nikon TU Plan Fluor, 10× with NA = 0.3; labeled L1 in Figure 3a), a 750 nm short-pass filter (FESH0750, Thorlabs; SP), a broadband 90:10 beam splitter (BSN10R, Thorlabs; BS1), a linear polarizer (WP25M-VIS, Thorlabs; P1), a defocusing lens (AC254-400-A-ML, Thorlabs; L2), and a broadband 50:50 beamsplitter (AHF analysentechnik; BS2), the laser light was focused on the back focal plane of an objective (TU Plan Fluor, 50× with NA = 0.8, Nikon; L3) on an inverted optical microscope (Eclipse Ti−U, Nikon). The diffracted light was then collected in reflection from the OFS, which was placed in a real plane of the imaging system. The light passed through lenses L4–L7 (tube lens, Nikon; AC254-200-A-ML, Thorlabs; AC254-200-A-ML, Thorlabs; AC508-200-A-ML, Thorlabs, respectively), several mirrors, an iris, and another linear polarizer (WP25M-VIS, Thorlabs; P2) before detection by a digital camera (Zyla PLUS sCMOS, Andor), which sat in a Fourier plane of the imaging system. The iris was placed in a real plane with a diameter that only collected light from the OFS. With the polarizers (P1, P2), we could independently analyze *s*- and *p*-polarized light. To account for power fluctuations of the laser over time, a fraction of the light was measured by a power meter (PM100D, Thorlabs; PM). The reported optical wavelength was carefully calibrated in all measurements.

**Evaluation and Analysis of Diffraction Efficiencies**. The measured diffraction efficiencies were evaluated and analyzed with Matlab (Release 2022b). The efficiencies were determined by normalizing signals from an OFS with that from a flat Ag reference surface. Additional background images were acquired to account for dark counts. After this correction, the total counts for each diffraction order $m$ were evaluated by integrating the counts within a square region of the Fourier image around the expected diffraction peak (see Figure 3b–d). The corresponding diffraction efficiency $\eta_m$ was calculated by dividing the total counts for diffraction order $m$ from the sample ($C_{m,\,\text{sample}}$) by the total counts from the reference ($C_{\text{tot, ref}}$), according to:

$$\eta_m = \frac{C_{m,\,\text{sample}}}{C_{\text{tot, ref}}} \cdot \frac{P_{\text{ref}}}{P_{\text{sample}}}, \tag{1}$$

which also corrects for the laser output powers, $P_{\text{sample}}$ and $P_{\text{ref}}$, for the sample and reference measurements, respectively.

**FDTD Simulations.** The simulations of the diffraction efficiency into the $-1^{\text{st}}$, $0^{\text{th}}$, and $+1^{\text{st}}$ orders were performed using the FDTD method as implemented in Lumerical FDTD (2023 R2.2, Ansys). The OFS was modelled as an infinitely large Ag slab with a spatially varying height profile. Permittivity data was taken from McPeak *et al.*,[51] which deposited Ag under similar conditions. The permittivity values were fitted and interpolated with a generalized multi-coefficient model in the 300–1000 nm wavelength range.

To simulate light diffracting from the Ag surface, a broadband wavepacket (plane-wave source) was injected 0.8 $\mu$m above the base of the OFS, as depicted in Figure S11 in the Supporting Information. During the simulation, the diffracted light was collected by a frequency-domain power monitor 0.1 $\mu$m above the plane-wave source. The FDTD simulation domain was 1 $\mu$m × 1.5 $\mu$m in the $xz$ plane and one mesh point in the $y$ direction (orange box in Figure S11 in the Supporting Information). By using periodic boundary conditions in $x$ and $y$, the simulated OFS was assumed to be infinitely large. Light could leave the simulation domain in $z$ *via* the perfectly matched layers (PMLs), which absorb the reflected and transmitted light and prevent back reflections from returning into the simulation domain. Mesh sizes of 5 and 0.5 nm were used for the simulations with *s*- and *p*-polarized light, respectively, which followed from convergence tests.

The diffraction efficiency was computed using the *grating* command in Lumerical. This function returns the normalized intensity of the diffraction orders in the far field, which are computed from the electric field recorded at the frequency-domain power monitor. To account for losses (*e.g.*, due to absorption in Ag), we corrected the normalized diffraction intensity from the *grating* command by the fraction of light that reached the monitor to correct for losses. For the comparison with experiment, the diffraction efficiency was determined by dividing the simulated reflection coefficient from the OFS by that from a flat Ag surface at each wavelength.

ASSOCIATED CONTENT

**Supporting Information.**

Detailed discussion of the different scalar diffraction models based on scalar diffraction theory and Fourier optics (Section S1). Additional figures (Section S2) and table (Section S3) discussing the schematics of the diffraction process, the different diffraction regimes, complementary data on the models, simulations, and experiments for the single- and double-sinusoidal OFSs (for *s*- and *p*- polarization), supplementary dark-field and SEM images, the FDTD simulation domain, and the design parameters for the OFSs in Ag.

AUTHOR INFORMATION

**Corresponding Author**

*Email: dnorris@ethz.ch.

**ORCID**

Yannik M. Glauser: 0000-0002-5362-0102

J. J. Erik Maris: 0000-0003-2591-4864

Raphael Brechbühler: 0000-0001-7498-9729

Juri G. Crimmann: 0000-0002-0367-5172

Valentina G. De Rosa: 0000-0002-8403-1802

Daniel Petter: 0000-0003-2788-1138

Gabriel Nagamine: 0000-0002-4830-7357

Nolan Lassaline: 0000-0002-5854-3900

David J. Norris: 0000-0002-3765-0678

**Note**

The authors declare no competing financial interest.

ACKNOWLEDGMENTS

This project was funded by ETH Zurich. N.L. gratefully acknowledges funding from the Swiss National Science Foundation (Postdoc Mobility P500PT_211105) and the Villum Foundation (Villum Experiment 50355). We thank H. Niese, D. B. Seda, and S. Vonk for stimulating discussions and J. Chaaban and U. Drechsler for technical assistance.

# Figures

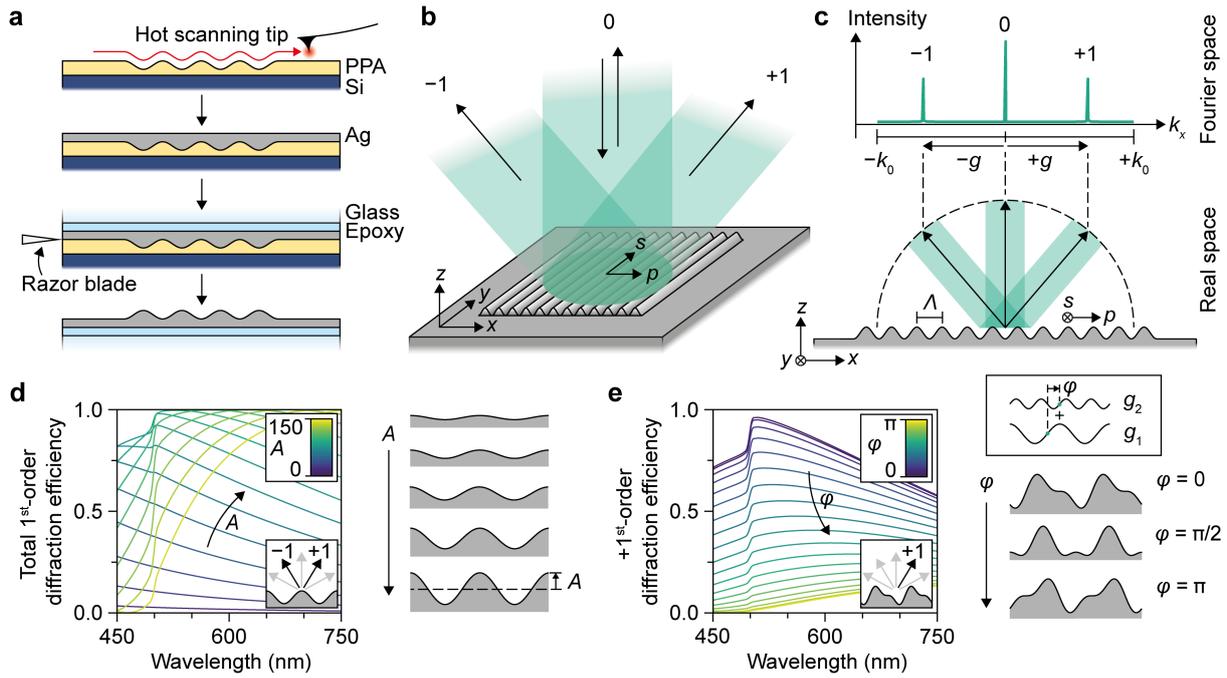

**Figure 1.** Key characteristics of the optical Fourier surface (OFS) platform. (a) Fabrication procedure of OFSs in Ag. (b) Schematic of light diffraction from an OFS in the form of a sinusoidal phase grating. The incoming light, either *s*- or *p*-polarized, is reflected and diffracted into the different diffraction orders. (c) The spatial angular frequency of the surface profile $g = 2\pi/\Lambda$ with spatial period $\Lambda$ in real space, represents the shift of the corresponding diffraction orders relative to each other in Fourier space. (d) Total 1$^{st}$-order diffraction efficiency for a single-sinusoidal OFS for amplitudes $A$ ranging from 0 to 150 nm. The period of the grating is fixed at $\Lambda$ = 1000 nm. (e) Diffraction efficiency of the +1$^{st}$ order for an OFS with two superposed sinusoids ($g_1$, $g_2$) with periods ($\Lambda_1$, $\Lambda_2$) = (1000 nm, 500 nm) and amplitudes ($A_1$, $A_2$) = (90 nm, 45 nm), respectively. The relative phase $\varphi$ between the two sinusoids is varied from 0 to $\pi$. The relative phase introduces a geometrical asymmetry in the system, which then leads to disproportionate diffraction into the −1$^{st}$ and +1$^{st}$ orders. The lower insets in (d,e) indicate the investigated diffraction order(s). The efficiency data in (d,e) is theoretically calculated based on a scalar diffraction model (Model 3).

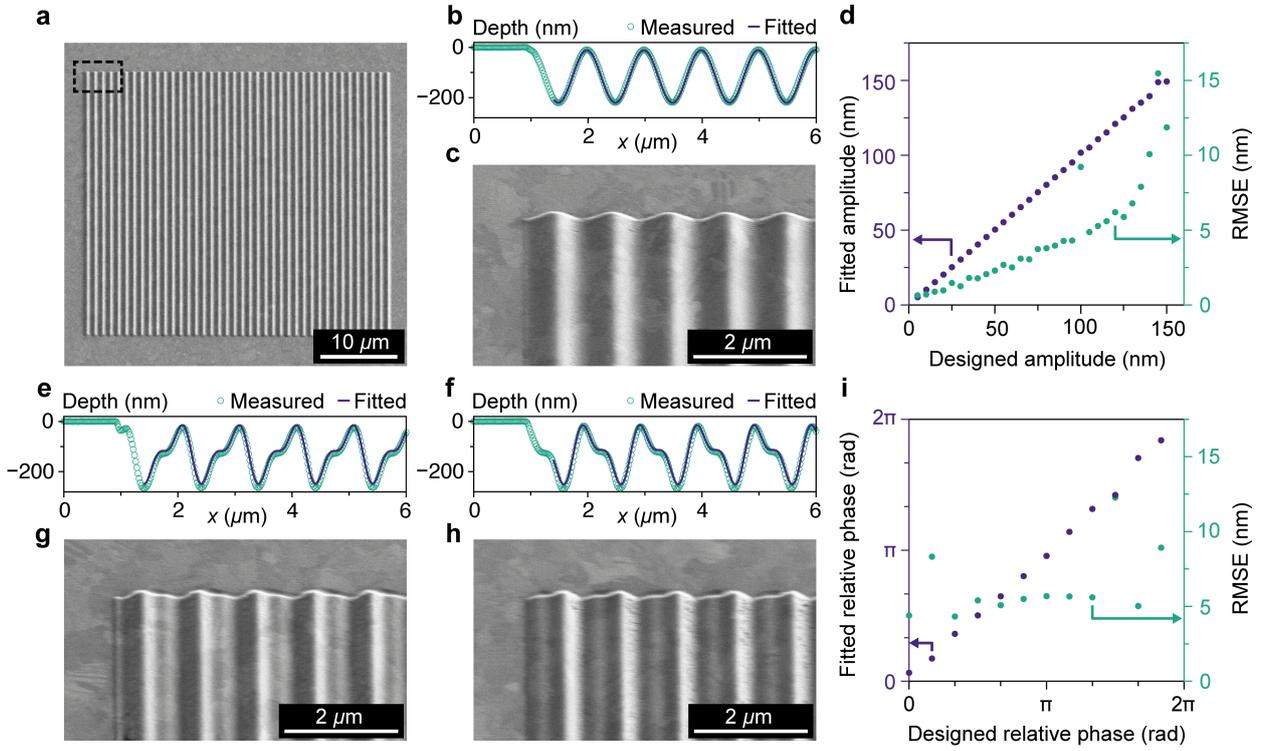

**Figure 2.** Surface topography characterization. (a) SEM image of a single-sinusoidal OFS in Ag with a period of 1000 nm and an amplitude of 105 nm. The lateral size of the grating is 40 $\mu$m × 40 $\mu$m. (b) Topography data of the patterned PPA film measured by the TSPL tool (circles) and the corresponding fit of the sinusoidal profile (line). The final structures in Ag are inverted and mirrored with respect to the structures in PPA due to template stripping. (c) Close-up of SEM image in (a), indicated by the black dashed rectangle. (d) Topography characterization of all 30 single-sinusoidal OFSs. The fitted amplitudes based on measurements are plotted as purple dots with respect to the left $y$ axis. (e,f) Measured and fitted surface profiles of double-sinusoidal OFSs in PPA with $\varphi = 0$ and $\varphi = \pi$, respectively. The sinusoids have periods of (1000 nm, 500 nm) with amplitudes of (90 nm, 45 nm), respectively. (g,h) SEM images of the top left corner of the double-sinusoidal OFSs in Ag whose profiles are depicted in (e,f), respectively. (i) Topography characterization of all 12 double-sinusoidal OFSs. The measured and fitted relative phases are plotted as purple dots with respect to the left $y$ axis. In (d,i), the root-mean-square error (RMSE) between the fitted function and the measured profile is depicted as green dots with respect to the right $y$ axis. SEM images were acquired under an angle of 30°.

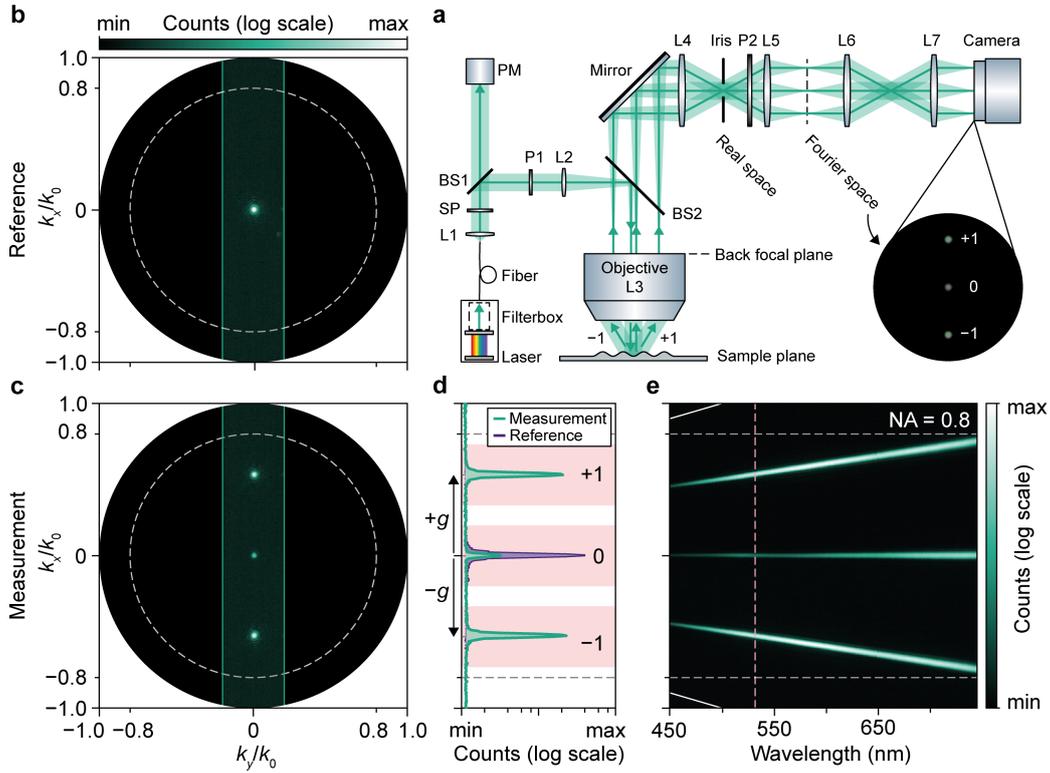

**Figure 3.** Diffraction-efficiency measurements and characterization. (a) Optical setup (Fourier microscope). Further details are presented in Methods. Measurements of the reflected and diffracted light in Fourier space for (b) a flat Ag surface (serving as reference) and (c) the single-sinusoidal OFS shown in Figure 2(a–c). The bright spots in the data represent the corresponding diffraction orders. For (b,c), the counts per pixel are depicted on the same logarithmic scale, located above (b). The wavelength was set to 532 nm. Only the area between the two vertical green lines has been measured to simplify the analysis. The circular black areas in Fourier space represent the light cone (NA = 1). The gray dashed circles denote the regions measurable by our objective (NA = 0.8). (d) Diffraction data shown in (b,c) integrated along $k_y$. (e) Diffraction data for different laser wavelengths. The vertical dashed pink line represents the measurement data in (c) at 532 nm. The solid white lines in (e) (which appear outside the NA, indicated by the gray horizontal dashed lines) denote the theoretical positions of the −2nd and +2nd orders, respectively. To extract diffraction efficiencies, the counts of all pixels within the pink area in (d) are additionally integrated along $k_x$ and normalized by the total reference counts.

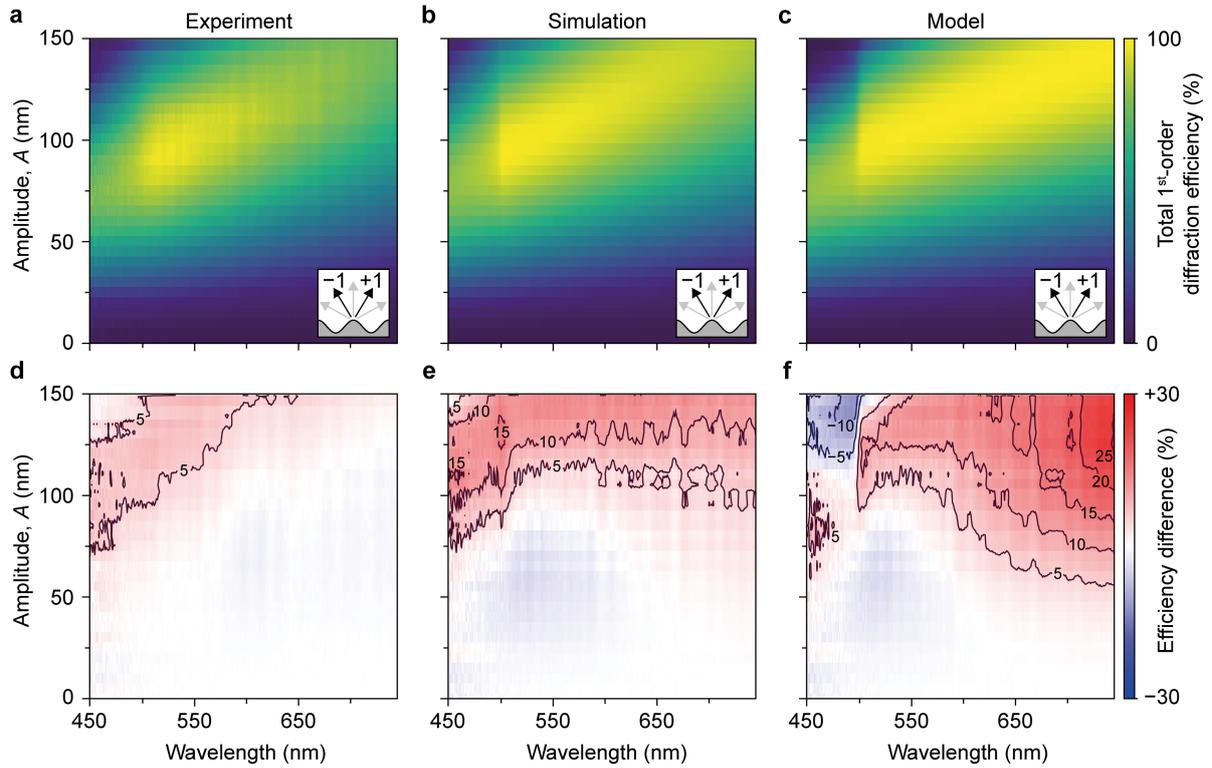

**Figure 4.** 1st-order diffraction efficiencies based on experiment, simulation, and model. (a–c) Total 1st-order diffraction efficiencies (sum of −1st and +1st orders) for single-sinusoidal OFSs with varying amplitudes for wavelengths ranging from 450 to 745 nm. The panels represent the collected data based on optical experiments, FDTD simulations, and a scalar diffraction model (Model 3), respectively. (d) Diffraction-efficiency difference between the −1st and +1st orders measured in the optical experiments. The differences in total 1st-order diffraction efficiency between (e) the simulation and experimental data and (f) a scalar diffraction model (Model 3) and the experimental data are shown. In (d−f), interpolated contour lines indicate differences in steps of 5%. On the $y$ axis, the fitted amplitudes are based on the surface-topography characterization (see Figure 2d). All measurements were conducted with $s$-polarized light ($p$-polarized results are shown in Figure S8 in the Supporting Information).

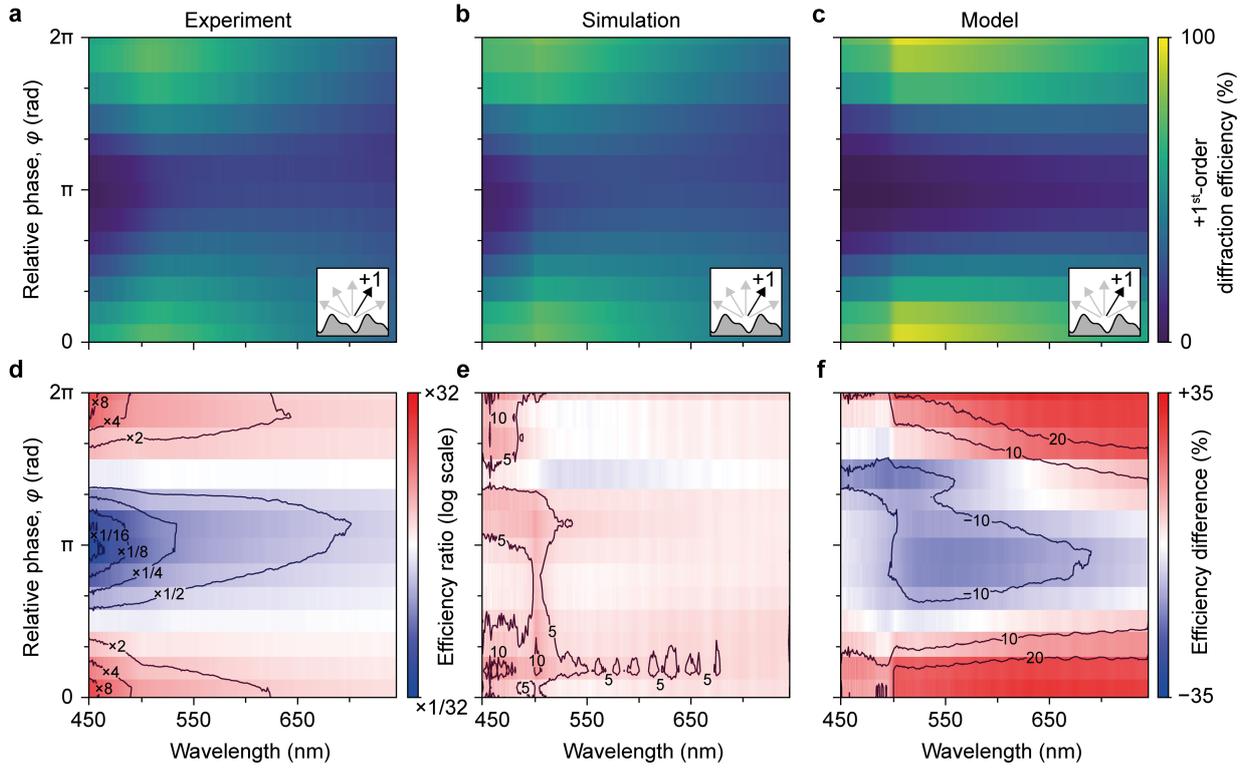

**Figure 5.** Asymmetric diffraction due to superposition of two sinusoids. (a–c) Diffraction efficiencies of the +1st order for double-sinusoidal OFSs with varying relative phase $\varphi$ for wavelengths ranging from 450 to 745 nm. The panels represent the collected data based on optical experiments, FDTD simulations, and a scalar diffraction model (Model 3), respectively. (d) Diffraction-efficiency ratios between the +1st and −1st orders measured in the optical experiments, quantifying the diffraction asymmetry. The ratio is plotted on a logarithmic scale with interpolated contour lines indicating ratios in powers of 2. The differences in the +1st-order diffraction efficiency between (e) the simulation and experimental data and (f) a scalar diffraction model (Model 3) and the experimental data are shown. (e,f) share the same linear scale, and interpolated contour lines indicate differences in steps of 5 and 10%, respectively. All measurements were conducted with *s*-polarized light.

**Table of Contents Graphic**

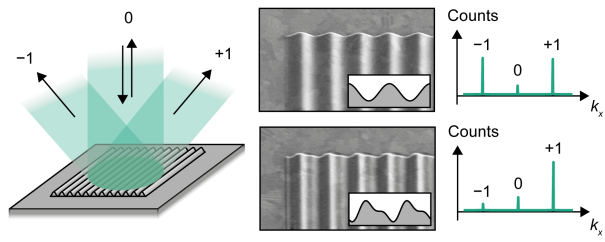

# Diffraction of Light from Optical Fourier Surfaces

*Yannik M. Glauser,[†] J. J. Erik Maris,[†] Raphael Brechbühler,[†] Juri G. Crimmann,[†] Valentina G. De Rosa,[†] Daniel Petter,[†] Gabriel Nagamine,[†] Nolan Lassaline,[†,§] and David J. Norris*[*,†]*

[†]Optical Materials Engineering Laboratory, Department of Mechanical and Process Engineering, ETH Zurich, 8092 Zurich, Switzerland

[§]Department of Physics, Technical University of Denmark, 2800 Kongens Lyngby, Denmark

**Corresponding Author**

*Email: dnorris@ethz.ch

# S1. SUPPLEMENTARY DISCUSSION

**Scalar Diffraction Theory and Fourier Optics.** The propagation of light as an electromagnetic wave is based on Maxwell's equations. Here, the absence of free charges and currents is considered. Because all components of the electric field **E** and magnetic field **H** satisfy the same identical wave equation, it is possible to summarize their behavior in a scalar wave equation:

$$\nabla^2 u(\mathbf{r}, t) - \frac{n^2}{c^2} \frac{\partial^2 u(\mathbf{r}, t)}{\partial t^2} = 0, \tag{S1}$$

where $u(\mathbf{r}, t)$ represents any of the scalar fields, which depend on position and time, $n$ is the refractive index of the dielectric medium, and $c$ is the speed of light in vacuum. Therefore, most diffraction theories use a scalar approach to discuss diffractive phenomena. Monochromatic waves can be described by

$$u(\mathbf{r}, t) = A(\mathbf{r}) \cos[\omega t - \phi(\mathbf{r})] \tag{S2}$$

with amplitude $A(\mathbf{r})$, phase $\phi(\mathbf{r})$, and angular frequency $\omega$. Similarly, the scalar field can be represented as the real part of a complex function by introducing the phasor $U(\mathbf{r})$, also known as the complex disturbance

$$u(\mathbf{r}, t) = \text{Re}\{U(\mathbf{r}) \exp[-i\omega t]\} \tag{S3}$$

with

$$U(\mathbf{r}) = A(\mathbf{r}) \exp[i\phi(\mathbf{r})]. \tag{S4}$$

Applying this definition to the scalar wave equation leads to the Helmholtz equation

$$(\nabla^2 + k^2) U(\mathbf{r}) = 0 \tag{S5}$$

where the harmonic time dependence of the monochromatic wave cancels out. The wavenumber $k$ is given by

$$k = \frac{n\omega}{c} = \frac{2\pi n}{\lambda}, \tag{S6}$$

where $\lambda$ denotes the wavelength of the monochromatic wave in vacuum. The simplest set of solutions satisfying the Helmholtz equation are waves of the form

$$U(\mathbf{r}) = \exp[i\mathbf{k} \cdot \mathbf{r}], \tag{S7}$$

described by the wavevector $\mathbf{k} = (k_x, k_y, k_z)$ with $|\mathbf{k}| = k$. Considering wave propagation in free space along the $z$ axis, the wavevector component $k_z$ can be defined as

$$k_z = \sqrt{k^2 - k_x^2 - k_y^2}, \qquad \text{Im}\{k_z\} \geq 0. \tag{S8}$$

Therefore, $k_z$ is either real or imaginary. For $k_x^2 + k_y^2 \leq k^2$, a solution as defined in eq S7 represents a propagating plane wave. In contrast, for $k_x^2 + k_y^2 > k^2$, eq S7 describes an exponentially decaying evanescent wave. Plane and evanescent waves are linearly independent with respect to each other and complete. Therefore, they form a basis for all potential solutions of the Helmholtz equation.

This result sets the foundation for Fourier optics, which describes the propagation of light waves based on the theory of Fourier analysis and linear systems. According to Fourier analysis, an arbitrary two-dimensional (2D) function $f(x, y)$ can be represented as a linear combination of harmonic functions with different in-plane wavevectors $(k_x, k_y)$. In this context, the wavevector refers to the specific spatial angular frequency of the harmonic function. A detailed discussion on Fourier optics can be found in Goodman,[S1] while a more compact overview is given in Saleh and Teich.[S2]

**Diffraction from Reflective Phase Gratings.** The concepts of scalar diffraction theory and Fourier optics are used to analytically describe the diffraction process of light from reflective sinusoidal phase gratings as well as the propagation of light within the optical setup. In our case, the reflective sinusoidal phase gratings are represented by optical Fourier surfaces (OFSs) in silver (Ag). The subsequent discussion is based on various works in the literature.[S1,S3-S5]

The disturbance $f(x, y)$ directly after reflection from a phase grating represents the input signal of our optical system (real space). It is determined by

$$f(x, y) = u_0 \cdot t(x, y) \tag{S9}$$

with an incoming reference wave $u_0$ and the corresponding grating transparency $t(x, y)$. For the following discussion, we assume an incoming plane reference wave, i.e., $u_0 = $ const. The transparency of a reflective phase grating is described by

$$t(x, y) = e^{i\Delta\varphi} = e^{i\frac{2\pi}{\lambda}OPD}. \tag{S10}$$

The optical path difference ($OPD$) represents the local difference in distance that the light must propagate relative to a flat reflective surface. It directly relates to the additional local phase $\Delta\varphi$ of the reflected light. $OPD$ is directly proportional to the surface profile $h(x, y)$ of the phase grating

$$OPD = -\gamma \cdot h(x, y) \tag{S11}$$

where $\gamma$ is a model specific parameter. The applied sign convention assumes a positive local phase for negative heights (longer propagation distance). Depending on the choice of the parameter $\gamma$, we can distinguish between paraxial and non-paraxial models:

(a) **Paraxial approach**. Paraxial scalar diffraction models use the paraxial assumption of small incident angles $\theta_i$ and small diffraction angles $\theta_m$ to diffraction order $m$. It is assumed that light perpendicularly enters and exits the surface profile at the same position with respect to a reference plane (see Figure S2a). This leads to $OPD$ of minus twice (in and out) the surface profile $h(x, y)$, i.e.,

$$\gamma_{\text{paraxial}} = 2. \tag{S12}$$

(b) **Non-paraxial approach**. In non-paraxial diffraction models, the incident and diffraction angles are not limited to small values only. In this case, it is assumed that light enters and exits the surface profile still at the same position with respect to reference plane, but under the corresponding angles $\theta_i$ and $\theta_m$ (see Figure S2b). This leads to $OPD$ consisting of an incident and a diffraction path with

$$\gamma_{\text{non-paraxial}} = \cos\theta_i + \cos\theta_m. \tag{S13}$$

In the subsequent section, we will discuss the paraxial and non-paraxial diffraction models for the single- and double-sinusoidal OFSs in a $2f$-system in free space, which represents a simplified optical setup. First, we will investigate the diffraction for the single-sinusoidal case and apply the derived formulas for the more general case with two superposed sinusoids. Single-sinusoidal OFSs have a surface profile with a single Fourier component

$$h(x,y) = A \sin(gx) \tag{S14}$$

with amplitude $A$, angular spatial frequency $g = \frac{2\pi}{\Lambda}$, and surface period $\Lambda$. The investigated double-sinusoidal OFSs consist of two superposed sinusoids of the form

$$h(x,y) = A \sin(gx) + \frac{A}{2}\sin(2gx - \varphi), \tag{S15}$$

where the spatial frequency $2g$ represents the second harmonic of the fundamental harmonic $g$. The amplitude of the former is chosen to be half of the latter. The relative phase between the two sinusoids is defined by $\varphi$. The transparency $t(x,y)$ of a single-sinusoidal phase grating can be described by

$$t(x,y) = e^{-i\tilde{A}\sin(gx)} = \sum_{m=-\infty}^{\infty} J_m(-\tilde{A}) e^{imgx} \tag{S16}$$

with

$$\tilde{A} = \frac{2\pi}{\lambda} \cdot A \cdot \gamma = k_0 \cdot A \cdot \gamma, \tag{S17}$$

where we applied the Jacobi–Anger expansion. $\tilde{A}$ represents the total phase amplitude, which is determined by the amplitude of the surface profile $A$, the parameter $\gamma$, and the wavenumber $k_0$ of the incoming light. Continuing from eq S16, we will derive the diffraction efficiencies based on different approaches. Based on the concepts of Fourier optics using the Fresnel approximation, the disturbance $g(x,y)$ in the output plane (Fourier space) of a $2f$-system can be described by the Fourier transform of the disturbance $f(x,y)$ in the input plane (real space):

$$g(x,y) = \frac{1}{i\lambda f} \mathcal{F}[f(x,y)](k_x, k_y) \tag{S18}$$

$$= \frac{4\pi^2 u_0}{i\lambda f} \sum_{m=-\infty}^{\infty} J_m(-\tilde{A})\, \delta(k_x - mg, k_y) \tag{S19}$$

with

$$k_x = \frac{k_0 x}{f},\, k_y = \frac{k_0 y}{f}. \tag{S20}$$

The shifted Dirac delta functions in Fourier space represent the different diffraction orders $m$. The diffraction peaks of the different diffraction orders are usually broadened due to finite-size effects. Accordingly, we can determine the intensity distribution at the output plane by

$$I(x, y) = |U(x, y)|^2. \tag{S21}$$

In this case, the intensity $I(x, y)$ represents the power density per collecting surface, also referred to as irradiance. We use the assumption that the overlap of the different diffraction orders with respect to each other is negligible. Therefore, terms of mixed diffraction orders are dropped. The intensity of a specific order $m$ and, therefore, also the corresponding diffraction efficiency only depend on the prefactors $c_m$ with

$$c_m = J_m^{\,2}(-\tilde{A}) = J_m^{\,2}(\tilde{A}). \tag{S22}$$

All the other preceding terms are constant. Based on this consideration, the diffraction efficiency remains independent of the incoming reference wave as long as we can assume that the overlap of the different diffraction orders is negligible.

In general, the diffraction efficiency of order $m$ describes the ratio between the diffracted power of order $m$ with respect to the total incoming power of the light, following

$$\eta_m = \frac{P_m}{P_{\text{tot}}}. \tag{S23}$$

Depending on the applied scalar diffraction model the efficiencies are determined slightly differently.

(a) **No renormalization.** Under the paraxial approximation of small incident and diffraction angles, the efficiency can directly be determined by

$$\eta_m^{\text{no renorm}} = R \frac{c_m}{\sum_{n=-\infty}^{\infty} c_n} = R c_m \tag{S24}$$

with reflectance $R$ of the Ag surface. We applied the fact that $\sum_{n=-\infty}^{\infty} c_n = 1$.

(b) **Renormalization.** In the non-paraxial regime, allowing larger incident and diffraction angles, diffraction orders can lie outside the light cone. Such orders involve evanescent waves which do not propagate. In the calculation of the diffraction efficiency, we therefore only normalize by orders $m'$ which correspond to propagating modes inside the light cone:

$$\eta_m^{\text{renorm}} = R \frac{c_m}{\sum_{m'=-\infty}^{\infty} c_{m'}}. \tag{S25}$$

This corresponds to the renormalization of the diffraction efficiency compared to the paraxial approach (Rayleigh anomaly), following

$$\eta_m^{\text{renorm}} = K \eta_m^{\text{no renorm}} \tag{S26}$$

with renormalization factor

$$K = \frac{\sum_{n=-\infty}^{\infty} c_n}{\sum_{m'=-\infty}^{\infty} c_{m'}} = \frac{1}{\sum_{m'=-\infty}^{\infty} c_{m'}}. \tag{S27}$$

(c) **Diffracted radiance.** Based on considerations in Harvey et al.,[S4,S5] in addition to renormalization, the non-paraxial prefactors $c_m$ should be normalized by the cosine of the diffraction angle $\theta_m$. This leads to corrected prefactors

$$c_{m,L} = \frac{c_m}{\cos \theta_m} = \frac{J_m^2(\tilde{A})}{\cos \theta_m}. \tag{S28}$$

This accounts for the fact that diffracted radiance $L$ (not irradiance or intensity) is shift-invariant in the direction cosine space (further details in Harvey et al.[S4,S5]). The evaluation of the diffraction efficiencies $\eta_m$ based on the prefactors $c_{m,L}$ can be analogously performed with eq S25.

To determine the diffraction efficiencies, the same concepts apply for the more general case of two superposed sinusoids in eq S15. The grating transparency is described by

$$t(x,y) = e^{-i\left[\tilde{A}\sin(gx) + \frac{\tilde{\tilde{A}}}{2}\sin(2gx-\varphi)\right]} \tag{S29}$$

$$= \sum_{m=-\infty}^{\infty} \sum_{n=-\infty}^{\infty} J_m(-\tilde{A}) J_n\left(-\frac{\tilde{\tilde{A}}}{2}\right) e^{-in\varphi} e^{i(m+2n)gx}, \tag{S30}$$

where we applied the Jacobi–Anger expansion formula as in eq S16. Substituting the index pair $(m, n)$ with $(l, n)$ according to

$$l = m + 2n, \tag{S31}$$

the grating transparency leads to

$$t(x,y) = \sum_{l=-\infty}^{\infty} \left[\sum_{n=-\infty}^{\infty} J_{l-2n}(-\tilde{A}) J_n\left(-\frac{\tilde{\tilde{A}}}{2}\right) e^{-in\varphi}\right] e^{ilgx}. \tag{S32}$$

Based on this expression the prefactors $c_l$ can be directly determined as

$$c_l = \left|\sum_{n=-\infty}^{\infty} J_{l-2n}(-\tilde{A}) J_n\left(-\frac{\tilde{\tilde{A}}}{2}\right) e^{-in\varphi}\right|^2. \tag{S33}$$

They can be used to analogously calculate the diffraction efficiencies according to eqs S23–S28. In theory, the same considerations on diffraction intensities and diffraction efficiencies can be extended to surface profiles with more Fourier components.

**Different Scalar Diffraction Models.** In this work, four different scalar diffraction models are presented and discussed. They represent all reasonable combinations of the different variations discussed above. They differ in the choice of the parameter $\gamma$, the presence or absence of the renormalization factor $K$, and the consideration of diffracted radiance with an additional factor $\cos\theta_m$.

- **Model 1.** Paraxial approach without renormalization: $\gamma = \gamma_{\text{paraxial}}$ and $\eta_m = \eta_m^{\text{no renorm}}$. This model consistently applies the paraxial approximation. In this case, orders with small diffraction angles carry most of the intensity, making an additional efficiency renormalization with $K$ unnecessary. It represents the simplest model. However, it is the only investigated model that does not account for Rayleigh anomalies.

- **Model 2.** Paraxial approach with renormalization: $\gamma = \gamma_{\text{paraxial}}$ and $\eta_m = \eta_m^{\text{renorm}}$. When surface profiles with many sinusoidal components are considered, the diffraction intensity in the output plane (Fourier space) is not concentrated on a few specific diffraction orders, but rather describes a continuous diffraction pattern. In this case, the paraxial approach with $\gamma_{\text{paraxial}} = 2$ enables the conversion of an arbitrary surface profile directly into a phase map $\Delta\varphi$, which can be used to calculate the intensity pattern in the output plane. This is not possible with the non-paraxial approach where $\gamma_{\text{non-paraxial}} = \cos\theta_i + \cos\theta_m$, because it only works for specific diffraction angles $\theta_m$. Using the paraxial approach with renormalization represents a more general extension to Model 1, which is specifically useful for surface profiles with many Fourier components and non-paraxial diffraction angles. The renormalization accounts for Rayleigh anomalies.

- **Model 3.** Non-paraxial approach with renormalization: $\gamma = \gamma_{\text{non-paraxial}}$ and $\eta_m = \eta_m^{\text{renorm}}$. This model uses a non-paraxial approach, which makes it generally applicable also to larger incident and diffraction angles. Because we only consider single-sinusoidal and double-sinusoidal OFSs with predefined diffraction angles in this work, this model is expected to estimate the diffraction efficiencies the most accurately. Therefore, Model 3 was used in the main text as representative scalar diffraction model for the direct comparison with simulations and optical experiments.

- **Model 4.** Non-paraxial approach with renormalization and corrected for diffracted radiance: $\gamma = \gamma_{\text{non-paraxial}}$, $\eta_m = \eta_m^{\text{renorm}}$, and $c_{m,L} = \frac{c_m}{\cos\theta_m}$. This model should be the most precise one according to considerations of Harvey et al.[S4,S5] It is based on diffractive radiance. However, the additional cosine term leads to infinitely large values for $c_{m,L}$ as the diffraction order approaches the edge of the light cone ($\theta_m = \frac{\pi}{2}$). In this case, the diffraction efficiencies approach 100% when the corresponding diffraction order become evanescent, leading to even stronger efficiency discontinuities (Rayleigh anomalies). Because this effect appears to be unintuitive and could not be directly observed in the simulations or the optical experiments, Model 3 was used for the discussion in the main text.

All four models have been used to estimate the diffraction efficiencies for the single-sinusoidal and double-sinusoidal OFSs over a wavelength range from 450 to 750 nm. The data is summarized in Figures S3 and S4, respectively. Comparing the model data of Figures S3 and S4 with the results from the simulations and optical experiments in Figures 4, 5, S7, and S9, we observe that Model 3 appears to best describe the diffraction process from our OFSs. This is consistent with our theoretical considerations that Model 3 represents the most general approach (non-paraxial approach with renormalization), which leads to physical results for Rayleigh anomalies (in comparison to Model 4).

## S2. SUPPLEMENTARY FIGURES

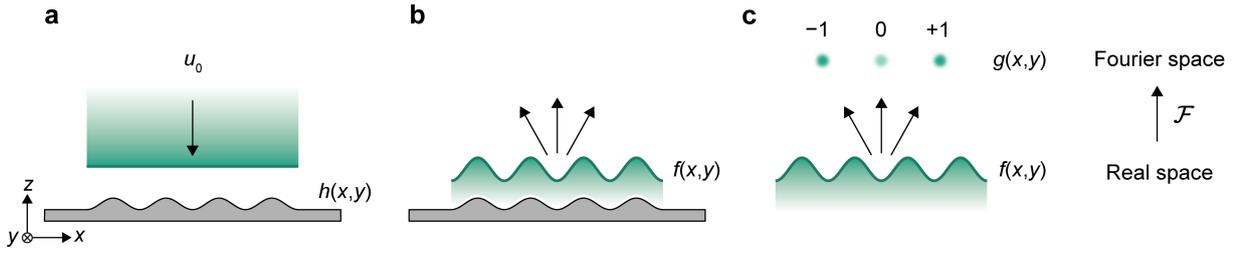

**Figure S1.** Schematic of the diffraction process. (a) Plane reference wave $u_0$ comes in at normal incidence with respect to the reflective optical Fourier surface (OFS). The OFS has a surface profile $h(x,y)$. (b) The wavefront directly after reflection. The surface profile introduces a phase modulation in the wavefront of light by creating optical path differences. (c) Reflected and diffracted wavefront $f(x,y)$ (real space) propagates into the far field, creating a diffraction pattern $g(x,y)$ (Fourier space). Under the Fraunhofer approximation,[S1] the relation between the diffraction signal in real space (input signal) and Fourier space (output signal) is represented by a Fourier transform. Similar to free-space propagation into the far field, a lens can perform a Fourier transform, but under the less constraining Fresnel approximation.[S1] In this case, the input and output plane must be placed at the front and back focal plane of the lens, respectively, representing a $2f$-system.

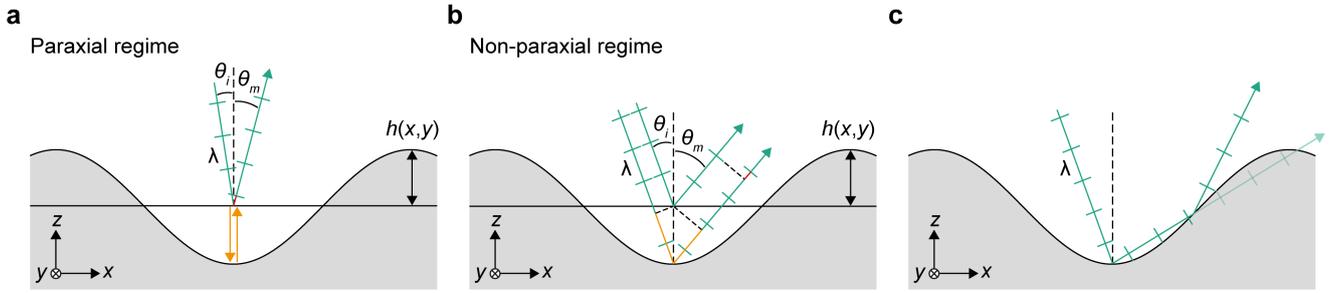

**Figure S2.** Diffraction regimes on a reflective phase grating. (a) The paraxial diffraction regime assumes small incident angles $\theta_i$ and small diffraction angles $\theta_m$ for the $m^{\text{th}}$ order. The optical path difference ($OPD$), introduced by a surface modulation $h(x,y)$ with respect to a reference plane, is represented by the vertical path length from the reference plane to the surface and back (orange), $OPD = -2h(x,y)$. (b) Non-paraxial diffraction regime accounts for larger incident and diffraction angles. $OPD$ is determined by the projection of the vertical path length on the direction of incidence and diffraction (orange), $OPD = -h(x,y)(\cos\theta_i + \cos\theta_m)$. In (a,b), the introduced phase modulation $\Delta\varphi$ (schematically highlighted in red) is calculated based on $OPD$ with respect to the wavelength $\lambda$ of light in free space, $\Delta\varphi = \frac{2\pi}{\lambda} \cdot OPD$. (c) Schematic configuration for which the assumptions of a shallow grating and small angles of incidence and diffraction do not hold. This can lead to diffraction of light into unwanted directions. The scalar diffraction models are not reliably applicable in this case.

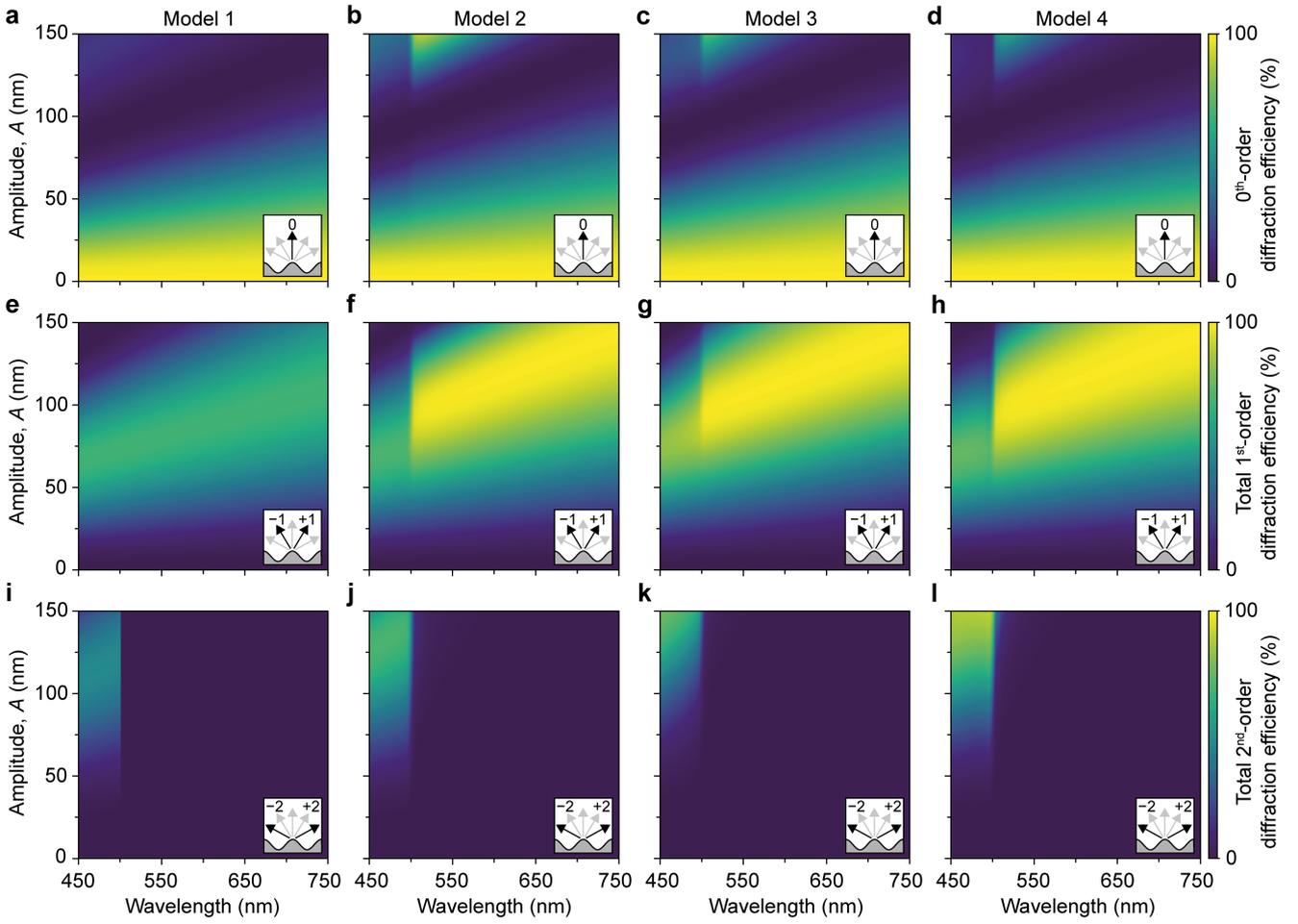

**Figure S3.** Comparison of scalar diffraction models for single-sinusoidal OFSs with a fixed period $\Lambda$ of 1000 nm. (a–d) $0^{th}$-order diffraction efficiency, (e–h) total $1^{st}$-order diffraction efficiency, and (e–h) total $2^{nd}$-order diffraction efficiency of single-sinusoidal OFSs. The design of the OFSs is the same as that depicted in Figure 1d of the main text. The wavelength of the light ranges from 450 to 750 nm. The data is based on different scalar diffraction models that are discussed in Section S1. Model 1 is applied for (a,e,i), Model 2 for (b,f,j), Model 3 for (c,g,k), and Model 4 for (d,h,l). The bottom right insets indicate the corresponding diffraction order(s) to which the data refers.

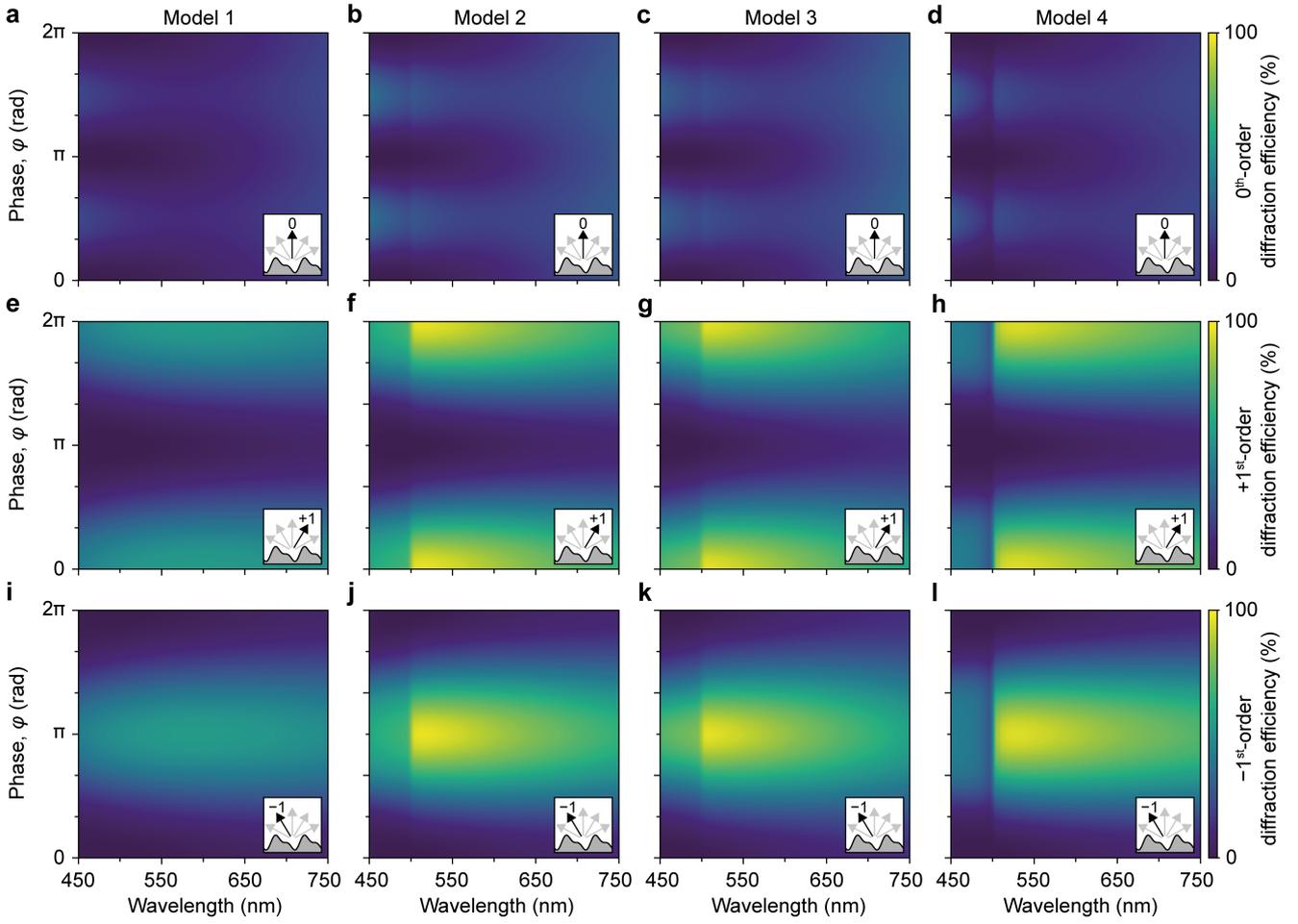

**Figure S4.** Comparison of scalar diffraction models for double-sinusoidal OFSs. (a–d) 0$^{th}$-order diffraction efficiency, (e–h) +1$^{st}$-order diffraction efficiency, and (e–h) −1$^{st}$-order diffraction efficiency of double-sinusoidal OFSs. The design of the OFSs is the same as that depicted in Figure 1e of the main text. The wavelength of the light ranges from 450 to 750 nm. The data is based on different scalar diffraction models that are discussed in Section S1. Model 1 is applied for (a,e,i), Model 2 for (b,f,j), Model 3 for (c,g,k), and Model 4 for (d,h,l). The bottom right insets indicate the corresponding diffraction order to which the data refers.

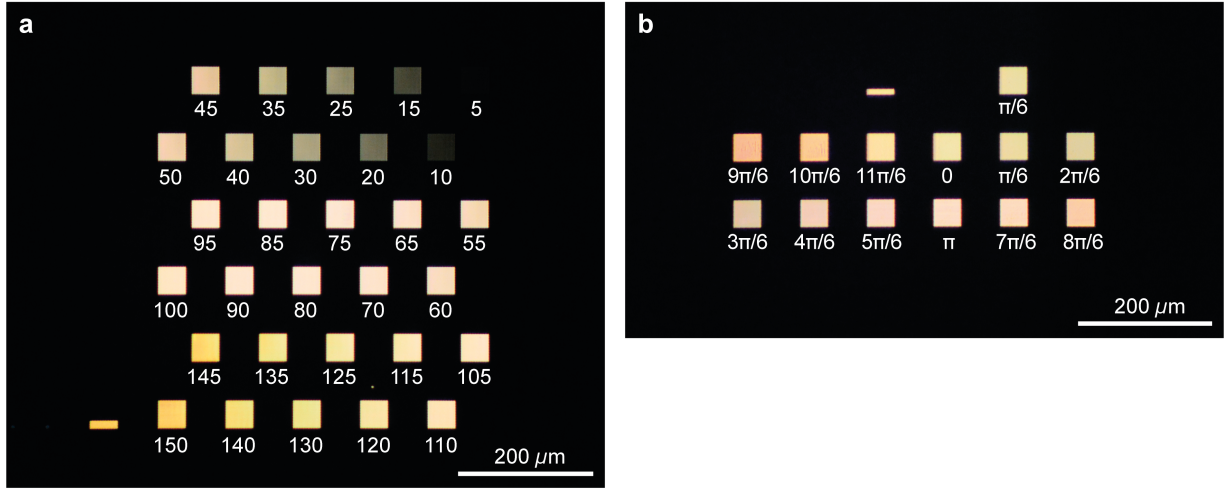

**Figure S5.** Dark-field images of OFSs. (a) Dark-field image of all single-sinusoidal OFSs in Ag that were analyzed in Figure 2d of the main text. The OFSs are indicated by their amplitude *A* in nm. They were patterned from left to right with decreasing amplitude. (b) Dark-field image of all double-sinusoidal OFSs in Ag that were analyzed in Figure 2i of the main text. The OFSs are indicated by the relative phase $\varphi$ in rad. The structure for $\varphi = \pi/6$ was patterned twice because the feedback of the thermal scanning-probe lithography (TSPL) tool initially did not work properly. The second OFS with the identical design, which was patterned on top of the initial one, was used later for the topological characterization and optical measurements. The smaller patterns in the bottom left corner of (a) and in the upper center of (b) were used as feedback calibration patterns.

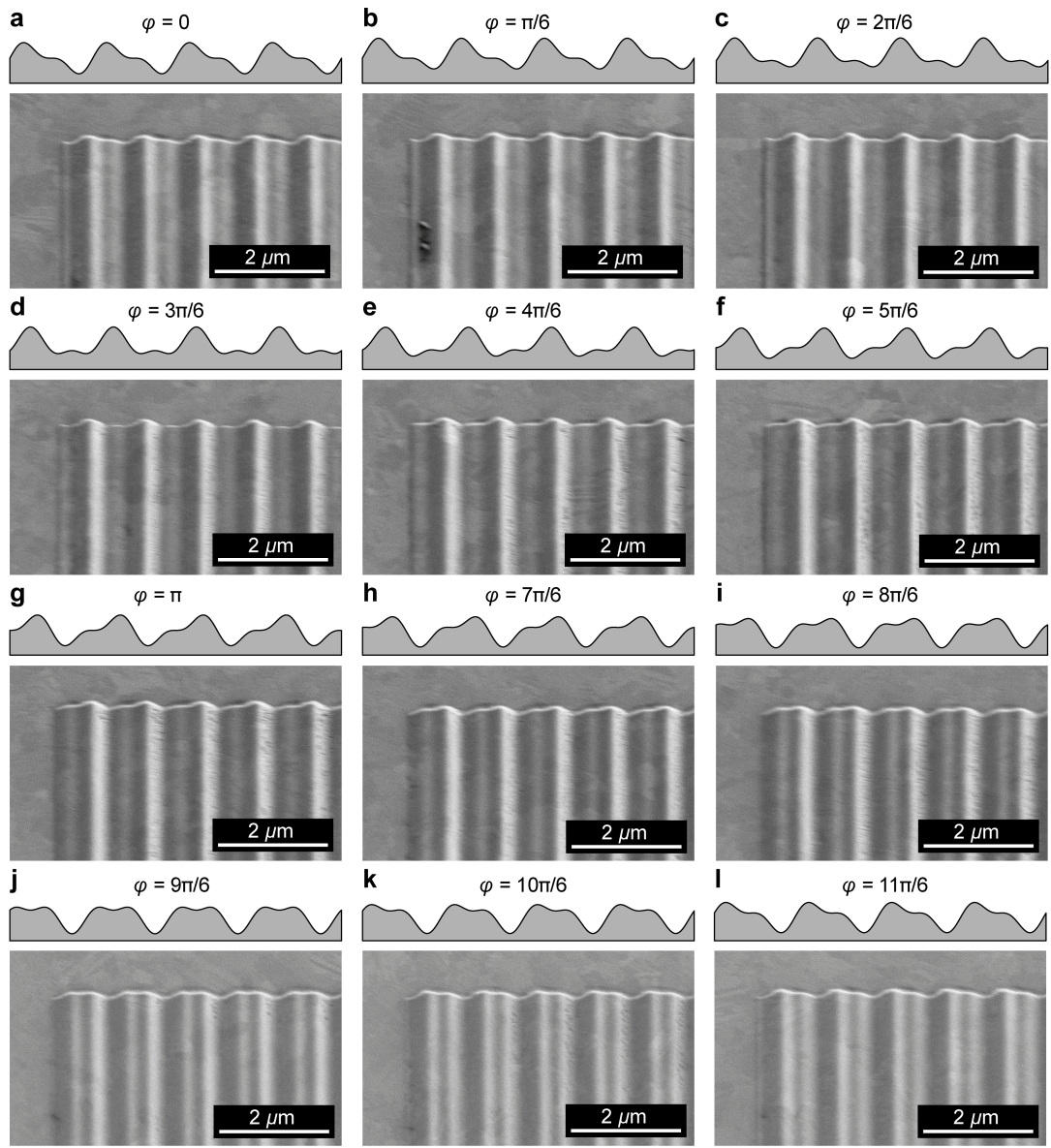

**Figure S6.** Schematic profiles and scanning-electron microscopy (SEM) images of the double-sinusoidal OFSs in Ag. (a–l) Top left corner of the double-sinusoidal OFSs in Ag, which are characterized in Figure 2i of the main text. The structures are tilted under an angle of 30°. The OFSs are defined by their relative phase $\varphi$. The corresponding schematic profiles are shown on top of the SEM images.

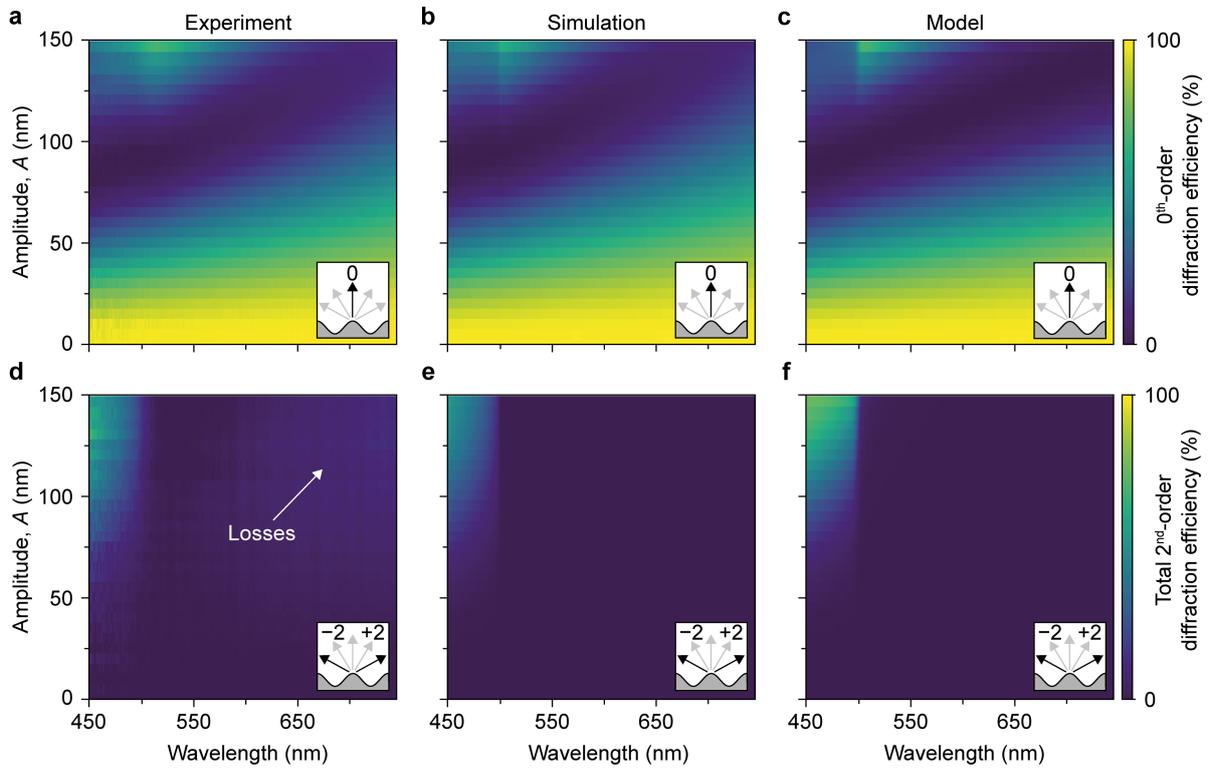

**Figure S7.** Diffraction efficiencies based on experiment, simulation, and model. Complementary data to Figure 4 of the main text. (a–c) 0$^{th}$-order diffraction efficiency and (d–f) total 2$^{nd}$-order diffraction efficiency for single-sinusoidal OFSs with varying amplitude for different wavelengths of light ranging from 450 to 745 nm. The panels represent the collected data based on optical experiments, finite-difference time-domain (FDTD) simulations, and a scalar diffraction model (Model 3), respectively (see Methods). In (d), due to the numerical aperture (NA) of our microscope objective (0.8), the total 2$^{nd}$-order diffraction efficiency could only be indirectly determined by using the uncollected photons with respect to the reference measurement. For wavelengths larger than 500 nm, the 2$^{nd}$ diffraction orders become evanescent and do not propagate anymore. These uncollected photons are therefore attributed to additional losses, *e.g.*, undesired diffraction of light into unmeasurable angles outside the NA (see Figure S2c) due to patterning imperfections or absorption due to PPA residues. Both effects are more likely for deeper structures. The FDTD simulations and the applied scalar diffraction model do not incorporate a limited NA due to an objective, but rather they assume NA = 1. (a–c) and (d–f) share the same linear scale, respectively. All measurements for this figure were conducted with *s*-polarized light.

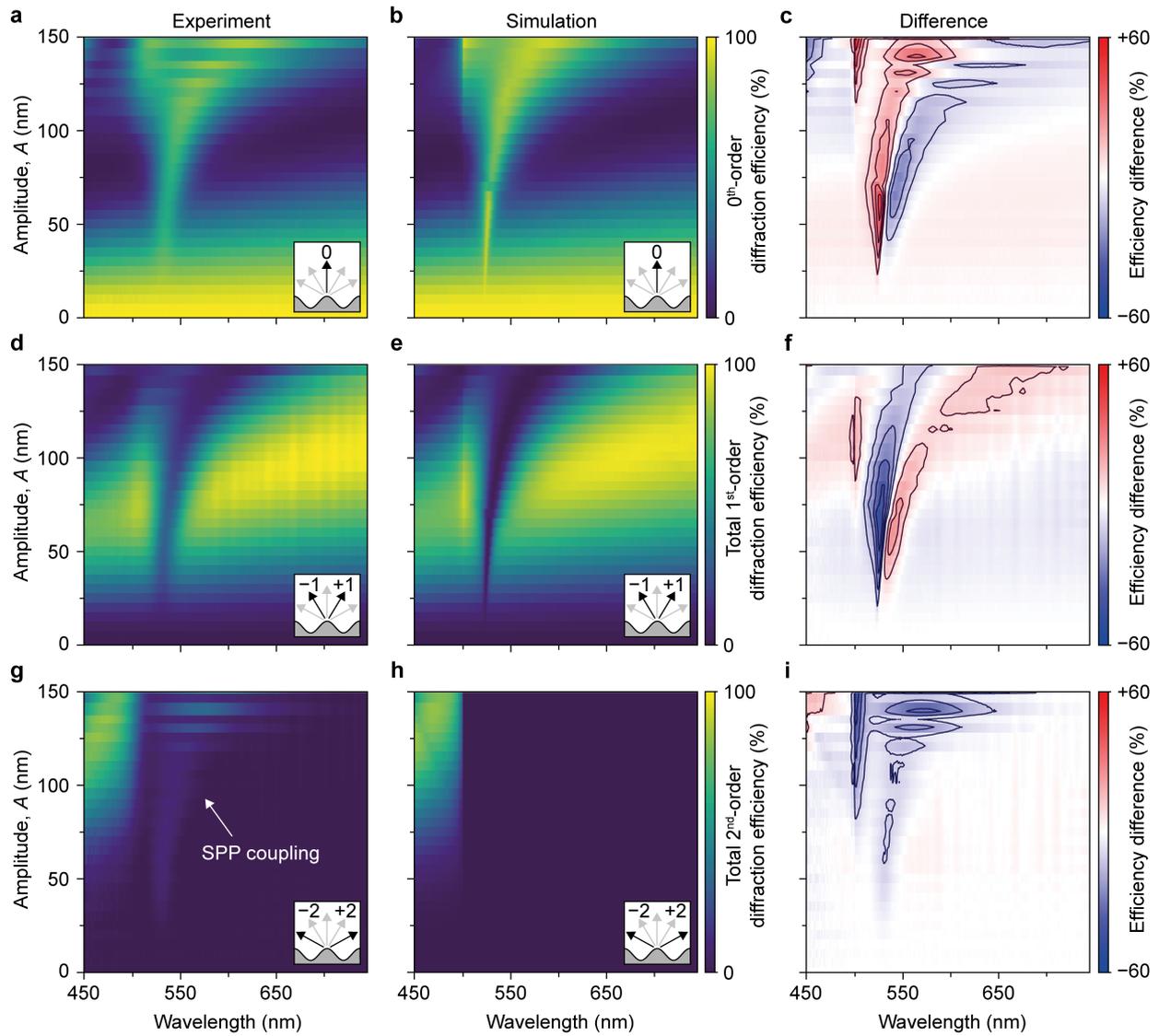

**Figure S8.** Diffraction efficiencies for *p*-polarized light. (a,b) 0th-order diffraction efficiency, (d,e) total 1st-order diffraction efficiency, and (g,h) total 2nd-order diffraction efficiency for single-sinusoidal OFSs with varying amplitude for different wavelengths of light ranging from 450 to 745 nm. The panels represent the collected data based on optical experiments and FDTD simulations, respectively. (c,f,i) show the corresponding differences between the results from the optical experiments and simulations, interpolated contour lines indicate steps of 10%. A feature is observed between 500 and 550 nm, which shows increased efficiency in (a,b,g) and reduced efficiency in (d,e) with respect to the corresponding data for *s*-polarized light in Figures 4 and S7. It corresponds to the coupling of *p*-polarized free-space photons to surface plasmon polaritons (SPPs) on the Ag/Air interface. The coupling feature appears to fan out and redshift toward larger amplitudes. Similar to Figure S7d, the total 2nd-order diffraction efficiency could only be determined indirectly *via* optical experiments, considering uncollected photons with respect to the reference measurement. Therefore, we observe SPP coupling losses in (g) for wavelengths longer than 500 nm.

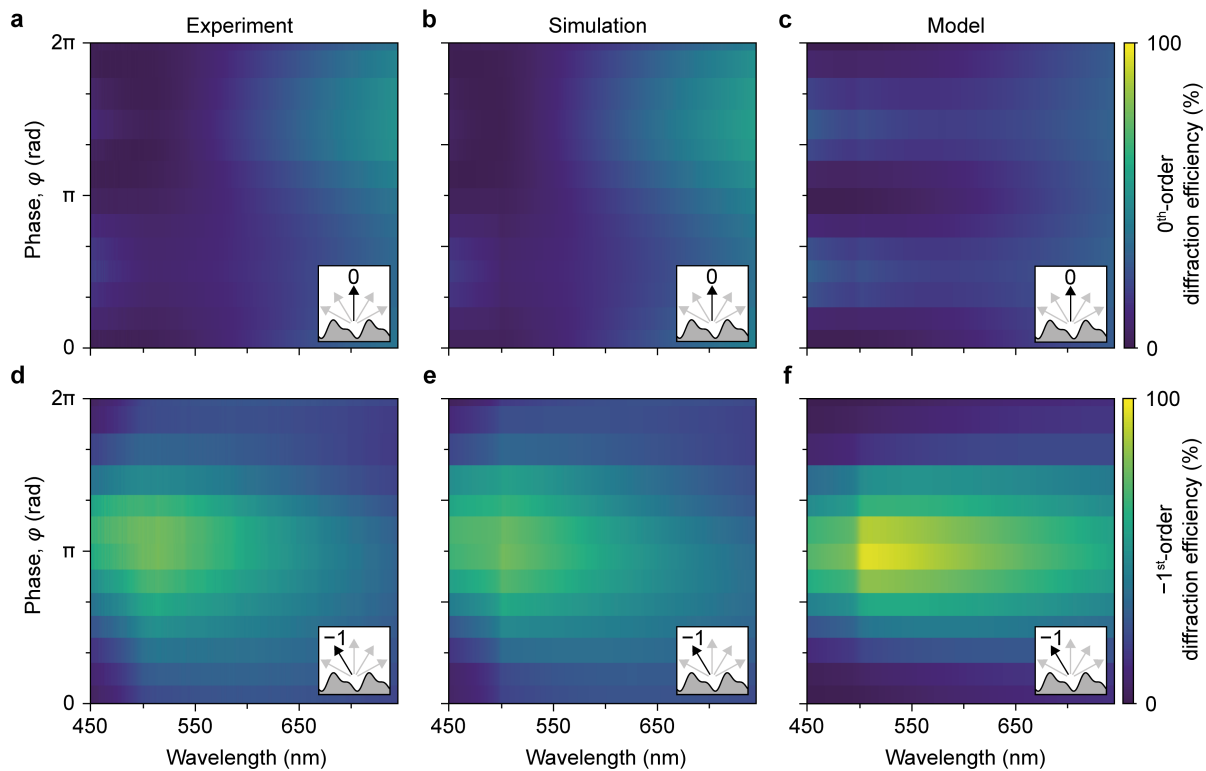

**Figure S9.** Asymmetric diffraction due to the superposition of two sinusoids. Complementary data to Figure 5 of the main text. (a–c) $0^{th}$-order diffraction efficiency and (d–f) $-1^{st}$-order diffraction efficiency for double-sinusoidal OFSs with varying relative phase for different wavelengths of light ranging from 450 to 745 nm. The panels represent the collected data based on optical experiments, FDTD simulations, and a scalar diffraction model (Model 3), respectively (see Methods). (a–c) and (d–f) share the same linear scale, respectively. All measurements for this figure were conducted with *s*-polarized light.

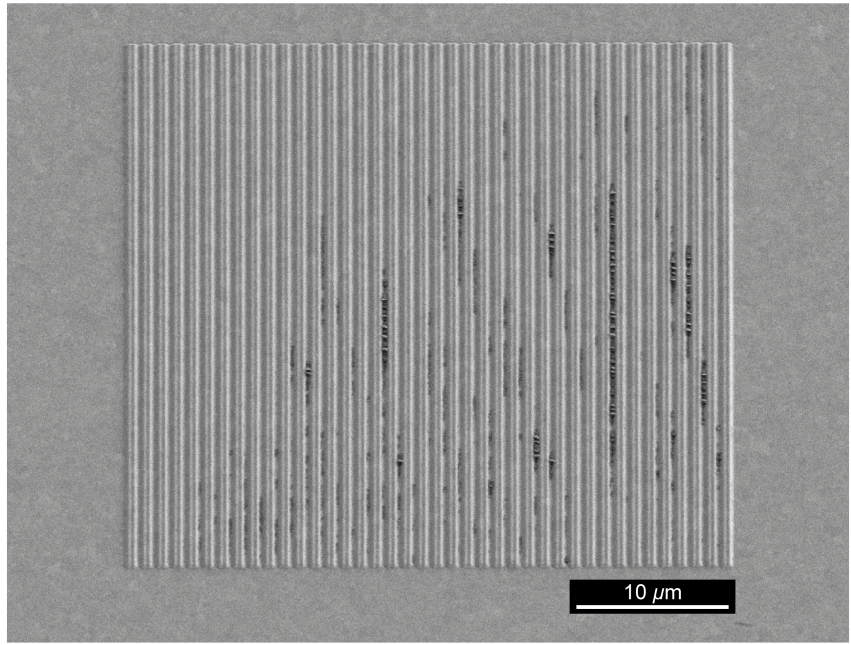

**Figure S10.** SEM image of a double-sinusoidal OFS in Ag with a relative phase $\varphi = 3\pi/2$. PPA residuals appear as dark spots on top of the Ag surface. They mostly occur in deeper topographic regions. The structure is tilted under an angle of 30°.

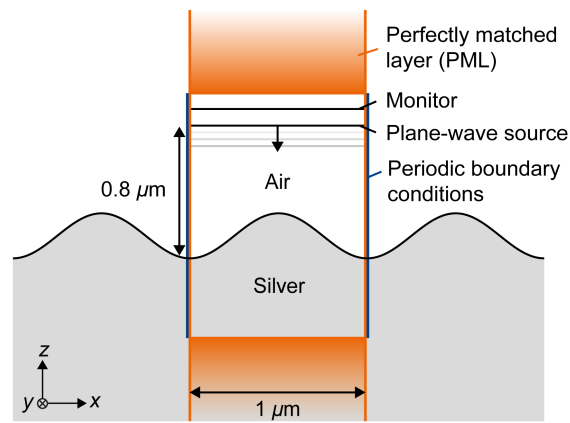

**Figure S11.** FDTD simulation configuration in Lumerical. The FDTD simulation domain is depicted in orange. Periodic boundary conditions along $x$ and $y$ were implemented with perfectly matched layers (PMLs) in $z$. A broadband plane wave was injected by a source above the surface structure. After reflection and diffraction, the light is collected by the monitor.

## S3. SUPPLEMENTARY TABLE

|  | Design parameters | | | | | Surface profile |
|---|---|---|---|---|---|---|
|  | $A_1$ (nm) | $A_2$ (nm) | $\Lambda_1$ (nm) | $\Lambda_2$ (nm) | $\varphi$ (rad) | $h(x,y)$ (nm) |
| **Single sinusoids** | 5–150 | – | 1000 | – | – | $A_1 \sin\left(\frac{2\pi}{\Lambda_1}x\right)$ |
| **Double sinusoids** | 90 | 45 | 1000 | 500 | 0–11π/6 | $A_1 \sin\left(\frac{2\pi}{\Lambda_1}x\right) + A_2 \sin\left(\frac{2\pi}{\Lambda_2}x - \varphi\right)$ |

**Table S1.** Design parameters for the final structures in Ag. The $z$ axis is chosen to be perpendicular to the sample plane ($xy$ plane), pointing away from the surface to form a right-handed coordinate system. The surface profiles for patterning PPA are accordingly adjusted to account for the process of template stripping.

# S4. SUPPLEMENTARY REFERENCES